\documentclass[twoside]{article}
\def\nkyear{200?}
\def\nkvol{??}
\def\nkno{?}

\def\firstpage{1}
\def\lastpage{20}
\def\correction{4}
\def\shorttitle{Cluster Partition Function and Invariants of 3-manifolds} 
\def\shortauthor{{\it M. Romo}} 
\usepackage b \baselineskip 14.5pt

\input{cam.doi}
\journalnumber{11401}
\DOImsnr{0001}
 \DOIyear{2007}

\title{\Large \bf \boldmath\ \\ Cluster Partition Function and Invariants of 3-Manifolds} 

\author{\large  Mauricio ROMO$^1$} 

\date{}

\begin{document}

\maketitle

\thispagestyle{first}
\renewcommand{\thefootnote}{\fnsymbol{footnote}}

\footnotetext{\hspace*{-5mm} \begin{tabular}{@{}r@{}p{13.4cm}@{}}
& Manuscript received  \\ 
$^1$ & School of Natural Sciences, Institute for Advanced Study, Princeton, NJ 08540, USA.\\
&{E-mail:mromoj-at-ias.edu} \\
\end{tabular}}

\renewcommand{\thefootnote}{\arabic{footnote}}

\begin{abstract} 
We review some recent developments in Chern-Simons theory on a hyperbolic 3-manifold $M$ with complex gauge group $G$. We focus on the case $G=SL(N,\mathbb{C})$ and with $M$ a knot complement. The main result presented in this note is the cluster partition function, a computational tool that uses cluster algebra techniques to evaluate the Chern-Simons path integral. We also review various applications and open questions regarding the cluster partition function and some of its relation with string theory.

\vskip 4.5mm

\nd \begin{tabular}{@{}l@{ }p{10.1cm}} {\bf Keywords } &
Chern-Simons theory, knots, cluster algebras
\end{tabular}

\nd {\bf 2000 MR Subject Classification } 
17B40, 17B50

\end{abstract}

\baselineskip 14pt

\setlength{\parindent}{1.5em}

\setcounter{section}{0}

\Section{Introduction} \label{section1}

In the recent years there has been a growing interest in the study of Chern-Simons (CS) theory with complex gauge group $G$, in particular when $G=SL(N,\mathbb{C})$. One of the main reasons is its appearance in the so called 3d-3d correspondence (see \cite{Dimofte:2014ija} for a review). This is a correspondence between supersymmetric 3d gauge theories and CS theory (which is a topological field theory) with complex gauge group. The correspondence arises from wrapping a certain class of 6-dimensional extended objects in M-theory, called M5-branes, on a 3-manifold $M$. This can be considered as part of a much broader line of research on the context of $\mathcal{N}=2$ supersymmetric quantum field theories. We refer the interested reader to the comprehensive review \cite{Teschner:2014oja}.\\

The main goal of this note is not to review the 3d-3d correspondence but to summarize some the properties and challenges present in CS theory with complex gauge group and, along with that, we will present in more detail a recently developed computational tool: the cluster partition function \cite{Gang:2015wya}. Formally, the partition function of CS theory with complex gauge group $G$ is given in the form of a path integral over $G$-connections $\mathcal{A}$:
\begin{eqnarray}\label{Path}
Z_{G}(M)=\int\mathcal{DA}\mathcal{D\overline{A}}e^{iS_{CS}[\mathcal{A},\overline{\mathcal{A}}]}
\end{eqnarray}
The main question is then, how can one make sense of $Z_{G}(M)$?. When $G$ is a compact Lie group, many techniques for computing $Z_{G}(M)$ and hence to give a definition of (\ref{Path}), have been developed since the pioneering work \cite{Witten:1988hf} connecting $Z_{G}(M)$ with invariants of 3-manifolds, but for noncompact and/or complex $G$ much less is understood. Foundational work on CS theory with complex gauge group can be found in \cite{Witten:1989ip} and, subsequent development in the lines that we will focus on here, in \cite{Gukov:2003na,Witten:2010cx}. One way to approach the problem in the case $\partial M\neq \emptyset$, is to interpret $Z_{G}(M)$ as a wavefunction. Geometric quantization of the restricted classical phase space of the theory\footnote{For the case of $G$ compact this was carried out in \cite{Witten:1988hf,Axelrod:1989xt,Elitzur:1989nr}.} associates a Hilbert space $\mathcal{H}_{\partial M}$ to $\partial M$. The CS partition function $Z_{G}(M)$ is then interpreted as a wavefunction in the following sense:
\begin{eqnarray}
Z_{G}(M)=\langle \mathcal{X},\Pi| \Psi \rangle \qquad \text{ \ for some \ } | \Psi \rangle\in \mathcal{H}_{\partial M}
\end{eqnarray}
where $\Pi$ represents a choice of polarization and $\mathcal{X}$, the 'position variables' (so, $\langle \mathcal{X},\Pi| \Psi \rangle$ is a function of $\mathcal{X}$ which are mutually commuting variables). The topology of $M$ (and possibly some extra data inherent to $M$) is what determines which vector $|\Psi\rangle$ should be chosen. When $M$ is a knot complement $M=S^{3}\setminus \mathcal{K}$, we have a toroidal boundary: $\partial M\cong T^{2}$ and the restricted classical phase space corresponds to flat $G$-bundles on $T^{2}$ and comes naturally equipped with a symplectic structure.\\

One of the most effective and well studied tools for computing partition functions $Z_{G}(M)=\langle \mathcal{X},\Pi| \Psi \rangle$ for the case of $M$ being a cusped 3-manifold and $G=SL(N,\mathbb{C})$ have been state-integral models. We will give a brief summary on development of these models in section \ref{sectquant}. The model we will present in detail, the cluster partition function, shares many properties with state-integral models and is expected to be equivalent in some cases.\\

The cluster partition function provides a way to define the CS path integral. In other words, it gives a prescription to compute a function $Z^{\mathrm{cluster}}_{G}(M)$ which should be interpreted as the wavefunction $Z_{G}(M)$. Our focus will be on the case of gauge group $G=SL(N,\mathbb{C})$ over a 3-manifold $M$ which corresponds to a (hyperbolic) knot complement on $S^{3}$. Moreover, we will actually see that $Z^{\mathrm{cluster}}_{G}(M)$ is more well suited for the case that $M$ can be obtained from a mapping torus construction. The cluster partition function was originally proposed in \cite{Terashima:2013fg} based on ideas of \cite{2011SIGMA...7..102K}, however important modifications were done in \cite{Gang:2015wya} to get it to the form we will present here. It is not immediately obvious that the function $Z^{\mathrm{cluster}}_{G}(M)$ we obtain is well defined as a nonperturbative invariant but we can propose perturbative topological invariants of $M$ starting from $Z^{\mathrm{cluster}}_{G}(M)$. We will look at this problem in more detail in section \ref{pertinvts}.\\

This note is organized as follows: in section \ref{CScpx} we will start reviewing classical aspects of the CS path integral for gauge group $G=SL(N,\mathbb{C})$ focusing on comparison with the case of compact gauge group, boundary conditions for the case of $M=S^{3}\setminus \mathcal{K}$ and some interpretations and results of the perturbative expansion of $Z_{G}(M)$. In section \ref{sectquant} we will review the canonical quantization of the boundary phase space for CS on $M=S^{3}\setminus \mathcal{K}$ and the subsequent interpretation of $Z_{G}(M)$ as a wavefunction. In section \ref{clustersec} we will come to the central theme of this note, we will review the derivation of $Z^{\mathrm{cluster}}_{G}(M)$ and explain its relation with $Z_{G}(M)$. For this we will start by reviewing the Fock-Goncharov construction of coordinates for the space of flat connections on a punctured Riemann surface $\Sigma$ and its quantization which defines a Hilbert space $\mathcal{H}_{\Sigma}$. Then, we will define $Z^{\mathrm{cluster}}_{G}(M)$ as the trace over $\mathcal{H}_{\Sigma}$ of an operator. We will see that our definition requires $M$ to be a mapping torus but its relation with cluster algebras (hence, its name) allows us to actually makes sense of $Z^{\mathrm{cluster}}_{Q,\mathbf{m}}$ as function of a general quiver $Q$ plus a sequence of mutations and permutations $\mathbf{m}$ acting on it. This will be an important point for section \ref{secappl} where we will review the applications of $Z^{\mathrm{cluster}}_{G}(M)$ and its general form $Z^{\mathrm{cluster}}_{Q,\mathbf{m}}$. Many of these applications are currently under study. Finally in the appendix we collect results about the quantum dilogarithm, a special function that plays a central role in the definition of $Z^{\mathrm{cluster}}_{Q,\mathbf{m}}$.

\Section{Chern-Simons theory with complex gauge group}\label{CScpx}

In this section we will review some basic facts about Chern-Simons (CS) theory with complex gauge group $G$. \footnote{Even though in this section we will make statements that hold for general $G$, in the rest of this note we will set $G=SL(N,\mathbb{C})$.} Consider a compact 3-manifold $M$, possibly with boundaries and a complex Lie group $G$. Fix a principal $G$-bundle $E_{G}\rightarrow M$ and consider a connection $\mathcal{A}$:
\begin{eqnarray}
\mathcal{A}\in \mathrm{Conn}(E_{G})
\end{eqnarray}
so, $\mathcal{A}$ can be seen as a $\mathfrak{g}$-valued 1-form on $M$, $\mathcal{A}\in \Omega^{1}(M,\mathfrak{g})$. We define the CS functional:
\begin{eqnarray}
CS[\mathcal{A}]:=\int_{M}\mathrm{Tr}\left(\mathcal{A}d\mathcal{A}+\frac{2}{3}\mathcal{A}\wedge \mathcal{A}\wedge\mathcal{A}\right)
\end{eqnarray}
and the CS action:
\begin{eqnarray}\label{CSaction}
S_{CS}:=\frac{t}{8\pi}CS[\mathcal{A}]+\frac{\tilde{t}}{8\pi}CS[\overline{\mathcal{A}}]\qquad t,\tilde{t}\in\mathbb{C}
\end{eqnarray}
Here $\overline{\mathcal{A}}=\mathcal{A}^{*}$ denotes the complex conjugate of $\mathcal{A}$. The coupling constants are conveniently written as
\begin{eqnarray}
t=k+is \qquad \tilde{t}=k-is\qquad k,s\in \mathbb{C}
\end{eqnarray}
and so (\ref{CSaction}) takes the form:
\begin{eqnarray}
S_{CS}:=\frac{k}{4\pi}\Re(CS[\mathcal{A}])-\frac{s}{4\pi}\Im(CS[\mathcal{A}])
\end{eqnarray}
We define the group of gauge transformations $\mathcal{G}$ along with its action on $\mathcal{A}$ as:
\begin{eqnarray}
\mathcal{G}:=\{ g:M\rightarrow G\}\qquad \mathcal{A}^{g}=g\mathcal{A}g^{-1}-dgg^{-1}
\end{eqnarray}
It is important to remark that $\mathcal{G}$ includes large gauge transformations\footnote{These are topologically nontrivial gauge transformations under which $\frac{1}{8\pi}CS[\mathcal{A}]$ shifts by $\pi\mathbb{Z}$.}. For $k\in\mathbb{Z}$ and $s\in \mathbb{C}$, then $e^{iS_{CS}}$ is gauge invariant \cite{Witten:1989ip}. Define
\begin{eqnarray}
\mathcal{Y}=\mathrm{Conn}(E_{G})/\mathcal{G}
\end{eqnarray}
If $\partial M\neq \emptyset$, we need to specify boundary conditions for $\mathcal{A}$, this means specify its behaviour at $\partial M$: $\mathcal{A}|_{\partial M}=\mathcal{A}_{b}$. We will make this more precise in section \ref{bdrysect}. The path integral associated to $(E_{G},M,\mathcal{A}_{b})$ is given by
\begin{eqnarray}\label{partfn}
Z_{G}(M)=\int_{\mathcal{Y}}\mathcal{DA}\mathcal{D\overline{A}}e^{iS_{CS}}=\int_{\mathcal{Y}}\mathcal{DA}\mathcal{D\overline{A}}e^{\frac{i}{4\pi}\left(k\Re(CS[\mathcal{A}])-s\Im(CS[\mathcal{A}])\right)}
\end{eqnarray}
where the boundary conditions should be imposed. If we restrict to $s\in \mathbb{R}$, then $Z_{G}(M)$ is an oscillatory integral. In such a case, $\tilde{t}=t^{*}$. However, other values of $s\in \mathbb{C}$ are also of interest. For instance, $Z_{G}(M)$ can also define an unitary theory for $s\in i\mathbb{R}$ \cite{Witten:1989ip}. Making sense of $Z_{G}(M)$ when $s\not\in \mathbb{R}$ is possible via analytic continuation \cite{Witten:2010cx}. We summarize the main cases treated in \cite{Witten:2010cx}:

\begin{itemize}
  \item \textbf{Compact gauge group}. Consider CS theory with gauge group a compact Lie group $H$. The construction is analogous as before, just changing $G$ to $H$, so the path integral for the connection $A\in\mathrm{Conn}(E_{H})$ is given by an integral over $\mathcal{Y}_{H}=\mathrm{Conn}(E_{H})/\mathcal{H}$:
      \begin{eqnarray}\label{pfcpct}
Z_{H}(M)=\int_{\mathcal{Y}_{H}}\mathcal{D}Ae^{\frac{i}{4\pi}kCS[A]}\qquad k\in\mathbb{Z}
\end{eqnarray}
Analytic continuation to $k\in\mathbb{C}$ is done in two steps \cite{Witten:1989ip,Witten:2010cx}. First change the domain of integration to be $\widehat{\mathcal{Y}}_{H}=\mathrm{Conn}(E_{H_{\mathbb{C}}})/\widehat{\mathcal{H}}_{\mathbb{C}}$ where the subindex $\mathbb{C}$ stands for complexification and $\widehat{\mathcal{H}}_{\mathbb{C}}$ are the topologically trivial gauge transformations $g:M\rightarrow H_{\mathbb{C}}$. Then define $Z_{H}(M)$ as the integration over a middle dimensional contour $\mathcal{C}\subset \widehat{\mathcal{Y}}_{H}$ such that $e^{\Re(ikCS[\mathcal{A}])}$ goes to $0$ along every asymptotic direction of $\mathcal{C}$. The contour $\mathcal{C}$ can be constructed by downward Morse flow using $\Re(ikCS[\mathcal{A}])$ as a Morse function. It is important to remark that $\mathcal{C}$ is not unique but one can impose extra conditions such as $Z_{H}(M)$ must coincide with the integration over $\mathcal{Y}_{H}\subset\widehat{\mathcal{Y}}_{H}$ when $k\in \mathbb{Z}$ and conditions on the behaviour for $k\rightarrow\infty$ to constraint the choices of $\mathcal{C}$. Another important point is that, because of the way $\mathcal{C}$ is chosen, it is not invariant under large gauge transformations. Indeed if one naively just change the domain of integration in (\ref{pfcpct}) to $\mathrm{Conn}(E_{H})/\widehat{\mathcal{H}}$, then (\ref{pfcpct}) will vanish if $k\not\in \mathbb{Z}$ \cite{Witten:2010cx}.
      \item \textbf{Compact gauge group with Wilson loop}. Consider now inserting a Wilson loop along a knot $\mathcal{K}\subset M$ in (\ref{pfcpct}). A Wilson loop is specified by $\mathcal{K}$ and an irreducible representation $R$ of $H$. Then,
          \begin{eqnarray}
Z_{H}(\mathcal{K})=\int_{\mathcal{Y}_{H}}\mathcal{D}Ae^{\frac{i}{4\pi}kCS[A]}\mathrm{Tr}_{R}Pe^{\oint_{\mathcal{K}}A}\qquad k\in\mathbb{Z}
\end{eqnarray}
We can write the holonomy operator $\mathrm{Tr}_{R}Pe^{\oint_{\mathcal{K}}A}$ as an integral over a quantum mechanical system, whose physical Hilbert space is identified with $R$, coupled to $A$ \cite{Beasley:2009mb,Witten:2010cx}:
 \begin{eqnarray}
\mathrm{Tr}_{R}Pe^{\oint_{\mathcal{K}}A}\sim \int_{\mathcal{U}}\mathcal{D}Ue^{I(A,u)}\qquad I(A,u)=\oint_{\mathcal{K}}(\Theta_{\alpha})_{m}\frac{d_{A}u^{m}}{d\tau}
\end{eqnarray}
where $\mathcal{U}$ is the space of maps $u:S^{1}\rightarrow H/T_{H}$,\footnote{Here $T_{H}$ denotes the maximal torus of $H$.} $\alpha$ is the highest weight of $R$, seen as an element of $T_{H}$ and $\Theta_{\alpha}=\mathrm{Tr}(\alpha g^{-1}dg)$ is a $1$-form in $H/T_{H}$. By writing the Wilson loop this way, we can write $Z_{H}(\mathcal{K})$ as an integral over $\mathcal{Y}_{H}\times \mathcal{U}$. Then, analytic continuation on $k$ follows the same steps as the case without Wilson loop but replacing the integration domain by a middle dimensional contour $\mathcal{C}\subset \widehat{\mathcal{Y}}_{H}\times \mathcal{U}_{\mathbb{C}}$ where $\mathcal{U}_{\mathbb{C}}$ is the space of maps $S^{1}\rightarrow H_{\mathbb{C}}/T_{H_{\mathbb{C}}}$.
  \item \textbf{Noncompact gauge group}. Here we will comment on the case of analytic continuation in $s$ of (\ref{partfn}). Renaming the connection $\overline{\mathcal{A}}\rightarrow \widetilde{\mathcal{A}}$ and taking $\widetilde{\mathcal{A}}\in \Omega^{1}(M,\mathfrak{g})$ as independent of $\mathcal{A}$ already makes (\ref{partfn}) non-invariant under large gauge transformations, as can be seen from its explicit form:
       \begin{eqnarray}\label{analyticcont}
e^{\frac{i}{8\pi}\left(tCS[\mathcal{A}]+\tilde{t}CS[\widetilde{\mathcal{A}}]\right)}=e^{\frac{i}{8\pi}\left(k(CS[\mathcal{A}]+CS[\widetilde{\mathcal{A}}])+is(CS[\mathcal{A}]-CS[\widetilde{\mathcal{A}}])\right)}
\end{eqnarray}
The functionals $CS[\mathcal{A}]/4\pi$ and $CS[\widetilde{\mathcal{A}}]/4\pi$ are defined modulo $2\pi$ but independent shifts will not leave (\ref{analyticcont}) invariant. So, integration over $\mathcal{Y}\times \mathcal{Y}$ is already ill defined. Only if both shifts are equal, (\ref{analyticcont}) is invariant. The solution proposed in \cite{Witten:2010cx} is to define the integral over $\widehat{\mathcal{Y}}_{G}$, the smallest covering of $\mathcal{Y}\times\mathcal{Y}$ on which (\ref{analyticcont}) is well defined, i.e. we quotient out only by the gauge transformations that leave (\ref{analyticcont}) invariant. Then, the analytic continuation of (\ref{partfn}) can be defined by integration over a middle dimensional contour  $\mathcal{C}\subset\widehat{\mathcal{Y}}_{G}$ which makes the partition function convergent. Let's remark on the relation with the case of compact gauge group \cite{Witten:2010cx}. Denote $H\subset G$ the unique compact subgroup satisfying $H_{\mathbb{C}}=G$. Denote the analytic continuation of CS partition function with group $H$ and level $k\in \mathbb{C}$ as
\begin{eqnarray}
Z_{H,\alpha}(M,k)=\int_{\mathcal{J}_{\alpha}\subset\widehat{\mathcal{Y}}_{H}}\mathcal{D}Ae^{\frac{i}{4\pi}kCS[A]}
\end{eqnarray}
where $\mathcal{J}_{\alpha}$ denotes a valid integration cycle. Since $\widehat{\mathcal{Y}}_{G}\subset \widehat{\mathcal{Y}}_{H}\times \widehat{\mathcal{Y}}_{H}$, and (\ref{analyticcont}) factorizes in a $\mathcal{A}$ and a $\widetilde{\mathcal{A}}$ dependent piece, then $\mathcal{C}$ can be factorized as $\mathcal{C}=\sum_{\alpha,\beta}m_{\alpha\beta}\mathcal{J}_{\alpha}\times \widetilde{\mathcal{J}}_{\beta}$ where $(\mathcal{J}_{\alpha},\widetilde{\mathcal{J}}_{\beta})$ are cycles in the basis of cycles under which $Z_{H,\alpha}(M,t/2-h^{\vee})$ and $Z_{H,\beta}(M,\tilde{t}/2-h^{\vee})$ are convergent.\footnote{Here $h^{\vee}$ denotes the dual Coxeter number of $H$ and the reason for this shift is explained in \cite{Witten:2010cx}.} Then we can write
\begin{eqnarray}
Z_{G}(M)&=&\int_{\mathcal{C}\subset\widehat{\mathcal{Y}}_{G}}\mathcal{DA}\mathcal{D\widetilde{\mathcal{A}}}e^{\frac{i}{8\pi}\left(tCS[\mathcal{A}]+\tilde{t}CS[\widetilde{\mathcal{A}}]\right)}\nonumber\\
&=&\sum_{\alpha,\beta}m_{\alpha\beta}Z_{H,\alpha}(M,t/2-h^{\vee})Z_{H,\beta}(M,\tilde{t}/2-h^{\vee})
\end{eqnarray}
\item \textbf{Noncompact gauge group with Wilson loop}. Now consider a finite dimensional representation $R$ of $H$ defined as before. Then we can lift this to a holomorphic or anti-holomorphic representation of $G=H_{\mathbb{C}}$. We choose a holomorphic lift. The partition function:
    \begin{eqnarray}
Z_{G}(\mathcal{K})=\int_{\mathcal{Y}}\mathcal{DA}\mathcal{D\overline{A}}e^{iS_{CS}}\mathrm{Tr}_{R}P e^{\oint_{\mathcal{K}}\mathcal{A}}
\end{eqnarray}
is already divergent, so we have no way to canonically define a contour of integration for it. The solution suggested in \cite{Witten:2010cx} is to define this path integral as:
\begin{eqnarray}
Z_{G}(\mathcal{K})=Z_{H}(\mathcal{K},t-h^{\vee})Z_{H}(M,\tilde{t}-h^{\vee})
\end{eqnarray}
where $Z_{H}(\mathcal{K},t-h^{\vee})$ is the analytic continuation of $Z_{H}$ with the insertion of a Wilson loop, defined as before, with just a shift of the level. So, in this case the partition $Z_{G}(\mathcal{K})$ completely factorizes. At the perturbative level, in $\frac{1}{k}$ this is equivalent to the statement that the expectation value of Wilson loops in a holomorphic representation obtained by complexifying a real one doesn't contain any new information. This is stated in exercise 6.32 in \cite{BN95} and also remarked in \cite{Witten:2010cx} and \cite{Gukov:2003na}.

\end{itemize}

\Subsection{Boundary conditions for $G=SL(N,\mathbb{C})$ and $M=S^{3}\setminus \mathcal{K}$ and classical phase space}\label{bdrysect}

From now on we will focus on the case of $G=SL(N,\mathbb{C})$ and $M$ being a knot complement in $S^{3}$:\footnote{Given a knot $\mathcal{K}$ embedded in $S^{3}$, set $N(\mathcal{K})$ a tubular neighborhood of $\mathcal{K}$ then $M=S^{3}\setminus \mathcal{K}$ is the compact manifold given by $S^{3}$ minus the interior of $N(\mathcal{K})$.}
\begin{eqnarray}
M=S^{3}\setminus \mathcal{K}
\end{eqnarray}
Then $M$ has a nonempty toroidal boundary:
\begin{eqnarray}
\partial M\cong T^{2}
\end{eqnarray}
Therefore, we distinguish two cycles, $\gamma_{m}$ and $\gamma_{l}$, in $\partial M$. The first cycle is called meridian and is given by the cycle that is contractible in $N(\mathcal{K})$. $\gamma_{l}$ is called longitudinal and is given by the transversal cycle to $\gamma_{m}$ in $\partial M$, intersecting it at a single point.\\

Boundary conditions for $\mathcal{A}$ are given by specifying the conjugacy class of its holonomy along a cycle $a\gamma_{m}+b\gamma_{l}$ ($a,b\in\mathbb{Z}$). For us, it will be better to consider specifying the holonomy along $\gamma_{l}$. For example if we fix the holonomy around $\gamma_{l}$ to a generic element of $SL(N,\mathbb{C})$ (i.e. an element such that all its eigenvalues $\{l_{i}\}_{i=1}^{N}$ are distinct), we get
\begin{eqnarray}
&&\mathrm{Hol}_{\gamma_{l}}(\mathcal{A})\sim\left(
                                         \begin{array}{cccc}
                                           e^{L_{1}} &  &  &  \\
                                            & e^{L_{2}} &  &  \\
                                            &  & \ddots &  \\
                                            &  &  & e^{L_{N}} \\
                                         \end{array}
                                       \right)=\left(
                                         \begin{array}{cccc}
                                           l_{1} &  &  &  \\
                                            & l_{2} &  &  \\
                                            &  & \ddots &  \\
                                            &  &  & l_{N} \\
                                         \end{array}
                                       \right)\nonumber\\
                                       \nonumber\\
                                       && L_{1}+\ldots+L_{N}=0\qquad l_{1}\cdots l_{N}=1
\end{eqnarray}
where the $\sim$ relation means up to conjugation by $SL(N,\mathbb{C})$. The classical phase space is given by extremizing the CS action $S_{CS}[\mathcal{A},\overline{\mathcal{A}}]$ and imposing the boundary conditions. The extrema of $S_{CS}$ is given by flat connections ($\mathcal{F}=d\mathcal{A}+\mathcal{A}\wedge\mathcal{A}$):
\begin{eqnarray}
\mathcal{F}=\overline{\mathcal{F}}=0
\end{eqnarray}
modulo gauge transformations. Flat connections modulo gauge transformations are in one-to-one correspondence with representations of $\pi_{1}(M)$ into the gauge group:
\begin{eqnarray}
\mathcal{M}_{\mathrm{flat}}(M)=\mathrm{Hom}(\pi_{1}(M),SL(N,\mathbb{C}))/SL(N,\mathbb{C})
\end{eqnarray}
where $SL(N,\mathbb{C})$ acts on $\mathrm{Hom}(\pi_{1}(M),SL(N,\mathbb{C}))$ by conjugation. The space $\mathcal{M}_{\mathrm{flat}}(M)$, also called $SL(N,\mathbb{C})$-character variety of $M$, should be described as a Lagrangian submanifold\footnote{One way to motivate this, is that a semiclassical solution can be characterized by a Lagrangian submanifold in the context of geometric quantization \cite{Woodhouse:1992de}.} $\mathcal{L}_{M}$ of $\mathcal{M}_{\mathrm{flat}}(\partial M)$ \cite{Dimofte:2013iv}, the moduli space of flat connections in $\partial M$. $\mathcal{L}_{M}$ is Lagrangian w.r.t the Weil-Petersson symplectic form on $\mathcal{M}_{\mathrm{flat}}(\Sigma)$, given by
\begin{eqnarray}
\omega_{WP}=\int_{\partial M}\mathrm{Tr}(\delta\mathcal{A}\wedge \delta\mathcal{A})
\end{eqnarray}
Since, in our case, $\pi_{1}(\partial M)=\pi_{1}(T^{2})=\mathbb{Z}\oplus \mathbb{Z}$, spanned by $\gamma_{l}$ and $\gamma_{m}$, then $\mathcal{M}_{\mathrm{flat}}(\partial M)$ has a very simple description:
\begin{eqnarray}
\mathcal{M}_{\mathrm{flat}}(\partial M)=((\mathbb{C}^{*})^{N-1}\times (\mathbb{C}^{*})^{N-1})/\mathcal{W}_{N}
\end{eqnarray}
Here the factors $(\mathbb{C}^{*})^{N-1}$ are spanned by the (independent) eigenvalues $(l_{1},\ldots,l_{N-1})$ and $(m_{1},\ldots,m_{N-1})$ of $\mathrm{Hol}_{\gamma_{l}}(\mathcal{A})$ and $\mathrm{Hol}_{\gamma_{m}}(\mathcal{A})$ respectively. $\mathcal{W}_{N}$ is the remaining action of the Weyl subgroup of $SL(N,\mathbb{C})$. Then, $\mathcal{L}_{M}\subset \mathcal{M}_{\mathrm{flat}}(\partial M)$ is described by a set of polynomial equations \cite{Dimofte:2009yn}:
\begin{eqnarray}\label{eqsN}
A_{i}(l,m)=0\qquad i=1,\ldots,N-1
\end{eqnarray}
We should make a few remarks before closing this subsection. First, note that the equations describing $\mathcal{L}_{M}$ are holomorphic. This is because $\mathcal{M}_{\mathrm{flat}}(\partial M)$ admits a hyper-K\"ahler structure \cite{HitchinSelfDuality} and $\mathcal{L}_{M}$ is a holomorphic subvariety. For the case $N=2$, there is only one equation $A(l,m)$ and  it is known to correspond to the classical A-polynomial \cite{CooperApolynomial} of the knot \cite{Gukov:2003na}.

\Subsection{Perturbative results for Chern-Simons theory with complex gauge group}

Perturbation theory of CS with compact gauge group has been studied for many years, citing all the existent literature would be overwhelming, so, instead we point out the interested reader to the review \cite{Freed:2008jq}. For the case of complex gauge group, perturbative aspects of CS theory has been analyzed in \cite{Gukov:2003na}, \cite{Dimofte:2009yn} and \cite{Dimofte:2012qj,Dimofte:2015kkp} for the case $G=SL(2,\mathbb{C})$. One of the main interests in this case is the connection between the perturbative expansion with the volume conjecture \cite{1996q.alg.....1025K,MurakamiMurakami,2002math......3119M} and the generalized volume conjecture \cite{Gukov:2003na}. \footnote{For a review see \cite{2010arXiv1003.4808D}.} Define the constants:
\begin{eqnarray}\label{hhparam}
\hbar:=\frac{4\pi i}{t}\qquad \tilde{\hbar}:=\frac{4\pi i}{\tilde{t}}
\end{eqnarray}
Recall that for the knot complement we are interested in, we fix the boundary condition by fixing the holonomy of $\mathcal{A}$ around one of the peripheral cycles. Suppose we fix $\mathrm{Hol}_{\gamma_{l}}(\mathcal{A})$ as a boundary condition, then the eigenvalues of $\mathrm{Hol}_{\gamma_{m}}(\mathcal{A})$ are fixed via the eqs. (\ref{eqsN}). A finite number of solutions $\{\mathcal{A}^{(\alpha)}\}_{\alpha\in I}$ is expected, labeled by a finite set of points $I \subset \mathcal{M}_{\mathrm{flat}}(M)$. Therefore, perturbatively, we expect that the partition function takes the factorized form:
\begin{eqnarray}
Z_{G}(S^{3}\setminus \mathcal{K})=\sum_{\alpha,\bar{\alpha}\in I}m_{\alpha\bar{\alpha}}Z^{\mathrm{pert}}_{\alpha}(\{l_{i}\};\hbar)\overline{Z}^{\mathrm{pert}}_{\bar{\alpha}}(\{\bar{l}_{i}\};\tilde{\hbar})
\end{eqnarray}
here, each $Z^{\mathrm{pert}}_{\alpha}(\{l_{i}\};\hbar)$ (resp. $\overline{Z}^{\mathrm{pert}}_{\bar{\alpha}}(\{\bar{l}_{i}\};\tilde{\hbar})$) can be seen as perturbation series on $\hbar$, (resp. $\tilde{\hbar}$) of $Z_{G}(M)$ around a critical point $\alpha\in \mathcal{M}_{\mathrm{flat}}(M)$. More precisely, each expansion takes the generic form \cite{Dimofte:2009yn}
\begin{eqnarray}\label{pertexp}
Z^{(\alpha)}_{G}(M)=\exp\left( \frac{1}{\hbar}S^{(\alpha)}_{0}-\frac{\delta^{(\alpha)}}{2}\ln\hbar+\sum_{n=1}^{\infty}S^{(\alpha)}_{n}\hbar^{n-1}\right)
\end{eqnarray}
As a final remark, as we mentioned before, this expansion for $G=SL(2,\mathbb{C})$, \footnote{This is the statement of the generalized volume conjecture \cite{Gukov:2003na}.} coincides with the expansion of the $SU(2)$ colored Jones polynomial $J_{n}(\mathcal{K},q)$, $q=e^{\frac{2\pi i}{k}}=e^{2\hbar}$ around $n\rightarrow\infty$, $k\rightarrow\infty$ while keeping $u=\frac{k}{n}$ fixed (the parameter $u$ is associated with the holonomy eigenvalues that we fix as boundary conditions \cite{2010arXiv1003.4808D}). This expansion in $\hbar$ can also be obtained by topological recursion applied to the 'character variety of $\mathcal{K}$' \cite{Dijkgraaf:2010ur,Borot:2012cw}, that is, the Lagrangian $\mathcal{L}_{M}$ defined in the previous subsection.

\Section{Quantization of classical phase space}\label{sectquant}

In this section we will consider and review the problem of quantizing the classical phase space of CS theory with $G=SL(N,\mathbb{C})$. This can be done, for example, using the Hamiltonian formalism (canonical quantization) \cite{Witten:1988hf}. Put the theory on $M=\mathbb{R}\times \Sigma$ where $\Sigma$ is a Riemann surface possibly with boundaries/punctures.\\
On $\mathbb{R}\times \Sigma$ we fix the natural gauge $\mathcal{A}_{0}=0$, where $x^{0}$ is the $\mathbb{R}$ direction on $\mathbb{R}\times \Sigma$. Then, the CS action (\ref{CSaction}) takes the form:
\begin{eqnarray}\label{CSHam}
S_{CS}=\frac{t}{8\pi}\int dx^{0} \int_{\Sigma} d^{2}z\mathrm{Tr}\left(\epsilon^{ij}\mathcal{A}_{i}\dot{\mathcal{A}}_{j}\right)+\frac{\tilde{t}}{8\pi}\int dx^{0} \int_{\Sigma} d^{2}z\mathrm{Tr}\left(\epsilon^{ij}\overline{\mathcal{A}}_{i}\dot{\overline{\mathcal{A}}}_{j}\right)
\end{eqnarray}
where $\dot{\mathcal{A}}=\partial_{x^{0}}\mathcal{A}$. The classical phase space is given by solutions to the equations of motion derived from $S_{CS}$, this is the space of flat connections on $\Sigma$:
\begin{eqnarray}
\dot{\mathcal{A}}_{j}=\dot{\overline{\mathcal{A}}}_{j}=0\qquad \overline{\mathcal{F}}_{ij}=\mathcal{F}_{ij}=0
\end{eqnarray}
Moreover, from (\ref{CSHam}) we can derive the Poisson brackets:
\begin{eqnarray}\label{PBts}
\{\mathcal{A}^{a}_{j}(x),\mathcal{A}^{b}_{j}(y)\}&=&\frac{8\pi}{t}\delta^{ab}\epsilon_{ij}\delta(x-y)\nonumber\\
\{\overline{\mathcal{A}}^{a}_{j}(x),\overline{\mathcal{A}}^{b}_{j}(y)\}&=&\frac{8\pi}{\tilde{t}}\delta^{ab}\epsilon_{ij}\delta(x-y)
\end{eqnarray}
This provides the classical phase space with a symplectic structure. Consider the case at hand, so the boundary is a torus:
\begin{eqnarray}
\partial(S^{3}\setminus \mathcal{K} )\cong T^{2}
\end{eqnarray}
then, the symplectic form induced from the Poisson brackets can be explicitly computed in terms of the holonomies around the cycles $\gamma_{m}$, $\gamma_{l}$:
\begin{eqnarray}
\omega_{T^{2}}=\frac{1}{4\pi}(t\Omega+\tilde{t}\overline{\Omega})\qquad \Omega=\sum_{i}\frac{dl_{i}}{l_{i}}\wedge \frac{dm_{i}}{m_{i}}=\sum_{i}dL_{i}\wedge dM_{i}
\end{eqnarray}

by choosing a real polarization we can quantize this phase space and the holomorphic part of the partition function (with fixed boundary condition) should satisfy:
\begin{eqnarray}\label{ApolyN}
\widehat{A}_{i}(\hat{l},\hat{m})Z_{\alpha}=0\qquad i=1,\ldots N-1
\end{eqnarray}

as was shown in \cite{Gukov:2003na} and in \cite{Dimofte:2011gm}, after choosing a real polarization of $(\mathcal{M}_{\mathrm{flat}}(T^{2}),\omega_{T^{2}})$, the operator $\widehat{A}(\hat{l},\hat{m})$ corresponds to the quantization of the A-polynomial of the knot $\mathcal{K}$ \cite{CooperApolynomial}, however, the quantization of $A(l,m)$ and, more generally, of the system $A_{i}(l,m)$ \footnote{A recursive method to find the quantization of the system of polynomial equations $A_{i}(l,m)(i=1,\ldots,N-1)$ is proposed in \cite{Dimofte:2009yn}.} is a complicated problem in general and this is not a practical way to compute $Z_{\alpha}$'s. The computation of CS partition function can be made by the use of the so-called state-integral models. The basic idea behind state-integral models is to start from a triangulation $\{\Delta_{i}\}_{i=1}^{N_{T}}$ of $M$ and assign a wave function to each $\Delta_{i}$, which can be then glued together via  process called symplectic gluing. For $G=SL(2,\mathbb{C})$ and level $(k=1,s)$, a state-integral model for shaped triangulations was developed in \cite{HikamiHyperbolic1,HikamiHyperbolic2,2007JGP} and its perturbative invariants further studied in \cite{Dimofte:2009yn}. An alternative state-integral model for $G=SL(2,\mathbb{C})$ and level $(k=1,s)$ was proposed in \cite{Dimofte:2011gm} and its perturbative invariants further analyzed in \cite{Dimofte:2012qj}. Nonperturbative aspects of the previous cases were recently studied in \cite{Andersen:2011bt}. The case of level $(k=0,s)$ was developed in \cite{Dimofte:2011py}. For the case of $G=SL(N,\mathbb{C})$ and level $(k=1,s)$, a perturbative model was developed in \cite{Dimofte:2013iv}. Very recently, the case of arbitrary level $(k,s)$ has been studied in \cite{Dimofte:2014zga,2014arXiv1409.1208E}. In the next section we will focus on a particular model which is called the cluster partition function.

\Section{Cluster partition function for mapping cylinder/torus}\label{clustersec}

We came to the main section of this note. Here we will review the construction of the cluster partition function developed in \cite{Gang:2015wya}. We will motivate it by its relation to $Z_{G}(M)$ for $G=SL(N,\mathbb{C})$ at level $(k=1,s)$ and $M=S^{3}\setminus \mathcal{K}$, when $M$ is a mapping torus. However in \cite{Gang:2015wya} the case $(k=0,s)$ was also analyzed and more importantly, as we will see, the cluster partition function can be constructed starting from any cluster algebra associated to a quiver.

\Subsection{Fock-Goncharov coordinates and quantization of $\mathcal{M}_{\mathrm{flat}}(\Sigma,\vec{\rho})$}


The classical phase space we are interested in is the moduli space of $SL(N,\mathbb{C})$-flat connections on a punctured Riemann surface $\Sigma$ modulo gauge transformations with prescribed holonomy around the punctures. Suppose $\Sigma$ has $h$ boundaries and $\{S_{a}\}_{a=1}^{h}$ denotes the generators of $\pi_{1}(\Sigma)$ corresponding to boundary curves. Also, consider a set $\{\rho_{a}\}_{a=1}^{h}$ of $h$ conjugacy classes $\rho_{a}$ of $SL(N,\mathbb{C})$. Then, define:
\begin{eqnarray}
\mathcal{M}_{\mathrm{flat}}(\Sigma,\vec{\rho})=\left\{\rho\in \mathrm{Hom}(\pi_{1}(\Sigma),SL(N,\mathbb{C}))/SL(N,\mathbb{C}):\rho(S_{a})\sim\rho_{a}\right\}
\end{eqnarray}
The space $\bigcup_{\vec{\rho}}\mathcal{M}_{\mathrm{flat}}(\Sigma,\vec{\rho})$, is a subspace of the space of framed $PGL(N,\mathbb{C})$ flat connections on $\Sigma$, which we denote by $\mathcal{X}_{N}(\Sigma)$. From now on we will work with $\mathcal{X}_{N}(\Sigma)$, but the final answer is expected to lift to a function in $\bigcup_{\vec{\rho}}\mathcal{M}_{\mathrm{flat}}(\Sigma,\vec{\rho})$. A similar situation occurs in \cite{Dimofte:2014zga}. To define $\mathcal{X}_{N}(\Sigma)$ we need to consider $\mathcal{F}_{N}$, the space of complete flags in $\mathbb{C}^{N}$, then
\begin{eqnarray}
\mathcal{X}_{N}(\Sigma)=\frac{\left\{(\rho,F^{1},\ldots,F^{h})\in \mathrm{Hom}(\pi_{1}(M),PGL(N,\mathbb{C}))\times (\mathcal{F}_{N})^{h}:\rho(S_{a})\text{ \ stabilizes \ }F^{a}\right\}}{PGL(N,\mathbb{C})}
\end{eqnarray}
here $PGL(N,\mathbb{C})$ acts on $\mathcal{F}_{N}$ by left multiplication. The space $\mathcal{X}_{N}(\Sigma)$ is a ramified cover of $\mathcal{M}_{\mathrm{flat}}(\Sigma)$ and it has been extensively studied in \cite{FockGoncharovHigher}. One of the results in \cite{FockGoncharovHigher} is the existence of a birational morphism:
\begin{eqnarray}
\mathcal{X}_{N}(\Sigma)\rightarrow (\mathbb{C}^{*})^{N_{0}}\qquad N_{0}=-(N^{2}-1)\chi(\Sigma)
\end{eqnarray}
where $\chi(\Sigma)=2-2g-h$ is the Euler characteristic of $\Sigma$. Fock and Goncharov (FG) in \cite{FockGoncharovHigher} constructed this map explicitly, providing us with explicit coordinates on a Zariski open subset of $\mathcal{X}_{N}(\Sigma)$. We will review the necessary aspects that will suit our needs. Consider the case of $\Sigma$ a once punctured torus:
\begin{eqnarray}
\Sigma=\Sigma_{1,1}=T^{2}\setminus \{ \mathrm{pt.} \}
\end{eqnarray}
Then, $\mathcal{M}_{\mathrm{flat}}(\Sigma_{1,1},\rho)$ is determined only by specifying one holonomy $\rho$ around the puncture. For the once punctured torus the mapping class group is homeomorphic to $SL(2,\mathbb{Z})$:
\begin{eqnarray}
MCG(\Sigma_{1,1})\cong SL(2,\mathbb{Z})
\end{eqnarray}
The coordinates on $\mathcal{X}_{N}(\Sigma_{1,1})$ constructed in \cite{FockGoncharovHigher} can be encoded in a quiver diagram, that can be draw on top of an ideal triangulation of $\Sigma$ (this is a triangulation of $\Sigma$ where the vertices correspond to punctures), with $N_{0}:=N^{2}-1$ nodes and adjacency matrix $Q_{i,j}$. \footnote{For a recent concise review of the construction of this space and the quiver diagram see the Appendix A of \cite{Coman:2015lna}.} See Figure 1 for an example.

\begin{figure}[t]
\centering
\includegraphics[width=3in]{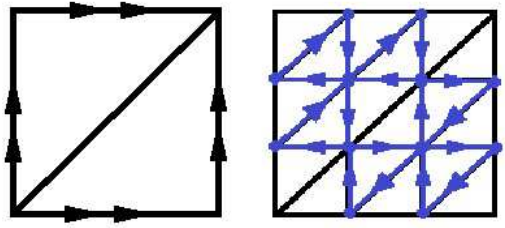}
\caption{On the left is an ideal triangulation of $\Sigma_{1,1}$ and the right is the FG quiver for $N=3$.}
\end{figure}

We will denote these coordinates by $\{y_{i}\}_{i=1}^{N_{0}}$. This construction has some remarkable features, one is that it equips $\mathcal{X}_{N}(\Sigma_{1,1})$ with a Poisson structure\footnote{The Poisson structure in the $y_{i}=\exp(Y_{i})$ coordinates is encoded in the matrix $Q_{i,j}$ by $\{Y_{i},Y_{j}\}\sim Q_{j,i}$.} and on the other hand, the action of $MCG(\Sigma)$ on these variables is given by cluster transformations generated by sequences of mutations of the quiver diagram (for a review of cluster algebras see \cite{FominZelevinsky}). A mutation at a node $k$ of $Q$ \footnote{In the following we will use $Q$ to denote the quiver and the adjacency matrix of the quiver as well.}corresponds to a change in the quiver's adjacency matrix given by:
\begin{eqnarray}
(\hat{\mu}_{k}Q)_{i,j}= \begin{cases}
    -Q_{i,j} & \text{if } i=k \text{ \ or \ } j=k\\
     Q_{i,j}+[Q_{i,k}]_{+}[Q_{k,i}]_{+}-[Q_{j,k}]_{+}[Q_{k,i}]_{+}& i,j\neq k
  \end{cases}
\end{eqnarray}
where $[x]_{+}=\mathrm{max}(x,0)$. The action of $\hat{\mu}_{k}$ on $Q$ is accompanied by an action on the $y_{i}$ variables:
\begin{eqnarray}
y'_{i}=\hat{\mu}_{k}(y_{i})= \begin{cases}
   y_{k}^{-1} & i=k\\
     y_{i}y_{k}^{[Q_{k,i}]_{+}}\left(1+y_{k}\right)^{-Q_{k,i}}& i\neq k
  \end{cases}
\end{eqnarray}
Because of the symplectic structure on $\mathcal{X}_{N}(\Sigma_{1,1})$ there exists a geometric quantization of this space, the $y$ variables are promoted to noncommutative ones:
\begin{eqnarray}
y_{i}\rightarrow\mathbf{y}_{i}
\end{eqnarray}

the $\mathbf{y}_{i}$'s satisfy commutations relations consistent with the symplectic form, forming a noncommutative algebra that we denote $A_{Q}$:
\begin{eqnarray}
A_{Q}=\{\mathbf{y}_{i=1,\ldots,N_{0}}|\mathbf{y}_{j}\mathbf{y}_{i}=q^{Q_{i,j}}\mathbf{y}_{i}\mathbf{y}_{j}\}\qquad q=e^{\hbar}
\end{eqnarray}

where $q$ is the quantization parameter. The logarithmic variables $\mathbf{Y}_{i}$ are defined by
\begin{eqnarray}
\mathbf{y}_{i}=e^{\mathbf{Y}_{i}}
\end{eqnarray}
they satisfy the relations:
\begin{eqnarray}\label{commY}
[\mathbf{Y}_{i},\mathbf{Y}_{j}]=\hbar Q_{j,i}\qquad \hbar=2\pi i b^{2}
\end{eqnarray}
where the parametrization $\mathbf{\hbar}=2\pi i b^{2}$ will prove useful in the subsequent sections. The quantization of the cluster transformations is given by:
\begin{eqnarray}
\mathbf{y}'_{i}=\hat{\mu}_{k}(\mathbf{y}_{i})= \begin{cases}
   \mathbf{y}_{k}^{-1} & i=k\\
     q^{\frac{1}{2}Q_{i,k}[Q_{k,i}]_{+}}\mathbf{y}_{i}\mathbf{y}_{k}^{[Q_{k,i}]_{+}}\prod_{m=1}^{|Q_{k,i}|}\left(1+q^{-\mathrm{sgn}(Q_{k,i})(m-\frac{1}{2})}\mathbf{y}_{k}\right)^{-\mathrm{sgn}(Q_{k,i})}\\
     =q^{\frac{1}{2}Q_{i,k}[-Q_{k,i}]_{+}}\mathbf{y}_{i}\mathbf{y}_{k}^{[-Q_{k,i}]_{+}}\prod_{m=1}^{|Q_{k,i}|}\left(1+q^{\mathrm{sgn}(Q_{k,i})(m-\frac{1}{2})}\mathbf{y}_{k}^{-1}\right)^{-\mathrm{sgn}(Q_{k,i})}& i\neq k
  \end{cases}
\end{eqnarray}
and this definition is such that, if we define $Q'=\hat{\mu}_{k}Q$, then the $\mathbf{y}'$ variables satisfy the relations:
\begin{eqnarray}
\mathbf{y}'_{j}\mathbf{y}'_{i}=q^{Q'_{i,j}}\mathbf{y}'_{i}\mathbf{y}'_{j}\qquad [\mathbf{Y}'_{i},\mathbf{Y}'_{j}]=\hbar Q'_{j,i}
\end{eqnarray}
by defining the following operators:
\begin{eqnarray}
\widehat{P}_{k}\mathbf{y}_{j}= \begin{cases}
    \mathbf{y}^{-1}_{k}\widehat{P}_{k} & \text{if } j=k\\
     q^{\frac{1}{2}Q_{i,k}[-Q_{k,i}]_{+}}\mathbf{y}_{i}\mathbf{y}^{[-Q_{k,i}]_{+}}_{k}\widehat{P}_{k}& \text{otherwise }
  \end{cases}
\end{eqnarray}
and
\begin{eqnarray}
e_{b}\left(\frac{-\mathbf{Y}_{k}}{2\pi b}\right)\mathbf{y}_{i}=\mathbf{y}_{i}\left(\prod_{m=1}^{|Q_{ki}|}(1+\mathbf{y}_{k}^{-1}q^{\mathrm{sgn}(Q_{k,i})(m-\frac{1}{2})})^{-\mathrm{sgn}(Q_{k,i})}\right)e_{b}\left(\frac{-\mathbf{Y}_{k}}{2\pi b}\right)
\end{eqnarray}
where $e_{b}(z)$ is the quantum dilogarithm function (see Appendix A for its definition and relevant properties), we can construct an explicit operator implementing the mutations:
\begin{eqnarray}
\hat{\mu}_{k}:=e_{b}\left(\frac{-\mathbf{Y}_{k}}{2\pi b}\right)\widehat{P}_{k}
\end{eqnarray}
So, this associates to $\mathcal{X}_{N}(\Sigma_{1,1})$ a noncommutative algebra and moreover, we can characterize the action of $MCG(\Sigma_{1,1})$ on $\mathcal{X}_{N}(\Sigma_{1,1})$ by cluster transformations. The precise sequences of mutations that correspond to generators of $MCG(\Sigma_{1,1})$ can be read in \cite{FockGoncharovHigher}, also see \cite{Gang:2015wya} for a review.

\Subsection{Cluster partition function at level $(k,s)=(1,s)$}

In this subsection we will present one of the main results of \cite{Gang:2015wya}: the cluster partition function. We will consider the case of CS level $(k=1,s)$ \footnote{In \cite{Gang:2015wya} the case $(k=0,s)$ is also analyzed, but here we will only focus on $k=1$.}. Let's first reparametrize the CS level in the following way:
\begin{eqnarray}\label{bparam}
s=-i\frac{1-b^{2}}{1+b^{2}}
\end{eqnarray}
where we consider $b\in\mathbb{C},|b|=1$, however analytic continuation to other values of $b$ such as $b\in \mathbb{R}$ is possible. The reason of this parametrization of $s$ comes from physics and the 3d-3d correspondence, see \cite{Cordova:2013cea,Hama:2011ea,Imamura:2011wg}, where the parameter $b$ has a very concrete interpretation in terms of the geometry of the M5-branes. In terms of (\ref{bparam}) the parameters (\ref{hhparam}) become:
\begin{eqnarray}\label{hh}
\hbar=2\pi i(1+b^{2})\qquad \tilde{\hbar}=2\pi i(1+b^{-2})
\end{eqnarray}
therefore we can define the parameters $q$, $\tilde{q}$ as
\begin{eqnarray}\label{qparam}
 e^{\hbar}=e^{2\pi i b^{2}}=q\qquad e^{\tilde{\hbar}}=e^{2\pi ib^{-2}}=\tilde{q}
\end{eqnarray}
From $\Sigma_{1,1}$ we can construct a knot complement through a mapping torus. For any punctured Riemann surface $\Sigma$, a mapping cylinder for an element $\varphi\in MCG(\Sigma)$, where $MCG(\Sigma)$ is the mapping class group of $\Sigma$, is given by:
\begin{eqnarray}
I_{\varphi}= (([0, 1] \times \Sigma) \amalg \Sigma)/\sim\qquad (0,x)\sim \varphi(x)
\end{eqnarray}
the mapping torus $M_{\varphi}$ is obtained by identifying both ends of $I_{\varphi}$:
\begin{eqnarray}
M_{\varphi}=(\Sigma\times S^{1})_{\varphi}:=\{(x,t)\in \Sigma\times [0,1]\}/\sim \qquad (x,0)\sim (\varphi(x),1)
\end{eqnarray}
If we take $\Sigma=\Sigma_{1,1}$, then $M_{\varphi}$ (with $\varphi\in SL(2,\mathbb{Z})$) can be identified with a hyperbolic knot complement whenever $|\mathrm{Tr}(\varphi)|>2$ \cite{Gueritaud}. As we saw in the previous section, the space $\mathcal{X}_{N}(\Sigma_{1,1})$ enjoys a quantization in terms of the Fock-Goncharov coordinates. We also saw that these coordinates only cover an open patch of $\mathcal{X}_{N}(\Sigma_{1,1})$ which we should identify with a subset of $\bigcup_{\rho}\mathcal{M}_{\mathrm{flat}}(\Sigma_{1,1},\rho)$. This is an important subject and we will return to it later. By now, let's work with the quantization of $\mathcal{X}_{N}(\Sigma_{1,1})$ given by the FG coordinates. For CS theory at levels $(1,s)$, the operators $\{\mathbf{y}_{i}\}_{i=0}^{N_{0}}$ with $N_{0}=N^{2}-1$ defined in the previous section satisfy the relations $A_{Q}$ with quantum parameter $q$ coinciding with (\ref{qparam}). Here we remark that the parametrization of $\hbar$ in terms of $b$ that we use in (\ref{commY}) differs from (\ref{hh}) by $2\pi i$. Since $q$ is the relevant parameter for the algebra $A_{Q}$, we can ignore this $2\pi i$ difference. In the following, everytime we write $\hbar$ we will mean
\begin{eqnarray}
\hbar=2\pi i b^{2}
\end{eqnarray}
The quiver $Q$ has $N^{2}-1$ vertices and a very simple form \cite{FockGoncharovHigher} and can be written as a tessellation of the ideal triangulation of $\Sigma_{1,1}$. See Section 4.5 of \cite{Gang:2015wya} for examples and applications of this, in the present context. The Poisson structure of the FG coordinates coincides with the holomorphic half of the Poisson structure that one can derive from Hamiltonian formalism \cite{Dimofte:2013iv}, putting CS theory on $\mathbb{R}\times \Sigma_{1,1}$. Indeed, given the set of $\mathbf{y}_{i}=e^{\mathbf{Y}_{i}}$ we can define the coordinates
\begin{eqnarray}
\tilde{\mathbf{y}}_{i}:=e^{\mathbf{Y}_{i}/b^{2}}
\end{eqnarray}
and they satisfy the relations:
\begin{eqnarray}
\tilde{\mathbf{y}}_{i}\tilde{\mathbf{y}}_{j}=\tilde{q}^{Q_{j,i}}\tilde{\mathbf{y}}_{j}\tilde{\mathbf{y}}_{i}\qquad \mathbf{y}_{i}\tilde{\mathbf{y}}_{j}=\tilde{\mathbf{y}}_{j}\mathbf{y}_{i}
\end{eqnarray}
therefore the $\tilde{\mathbf{y}}_{i}$ coordinates are identified with the anti-holomorphic side. \footnote{In this particular case, i.e., when $k=1$, the variable $Y_{i}$ can be parametrized as $Y_{i}=bK_{i}$ where $K_{i}$ is real, hence, for $|b|=1$, the conjugate $\overline{Y}_{i}=b^{-2}Y_{i}$. For more details on this see \cite{Dimofte:2014zga}This is why the case $k=1$ is actually very close to the situation of CS theory with gauge group $G=SL(N,\mathbb{R})$. For arbitrary level $k$ the situation is more complicated, as analyzed in \cite{Dimofte:2014zga}.}. The operators $\mathbf{y}$ can be acted upon by mutations $\hat{\mu}_{k}$ which will change their algebra of relations to $A_{Q'}$ and also we will consider permutations $\hat{\sigma}$, defined by a permutation matrix
\begin{eqnarray}
\sigma\in S_{N_{0}}\subset GL(N_{0},\mathbb{Z})
\end{eqnarray}
Operators $\hat{\sigma}$ will act on the $\mathbf{y}$ variables as
\begin{eqnarray}
\hat{\sigma}(Q)=\sigma\cdot Q\cdot\sigma^{T}\qquad \hat{\sigma}(\mathbf{y})_{i}=\sigma_{i,j}\mathbf{y}_{j}
\end{eqnarray}
After quantization, the operators $\mathbf{y}_{i}$ act on a Hilbert space that we will denote by $\mathcal{H}_{Q}$. Before describing $\mathcal{H}_{Q}$ more explicitly, let's define a function, in terms of a chain of $t_{f}$ permutations and mutations, which we denote simply by $\mathbf{m}$\footnote{In our convention we read the products from left to right.}:
\begin{eqnarray}
\mathbf{m}:=\hat{\mu}_{m_{0}}\hat{\sigma}(0)\cdots \hat{\mu}_{m_{t_{f}-1}}\hat{\sigma}(t_{f}-1)
\end{eqnarray}
and denote the quiver at 'time' $t$ by $Q(t)$ (and we denote $Q$ the original quiver):
\begin{eqnarray}
Q(t)=\hat{\sigma}(t-1)\hat{\mu}_{m_{t-1}}\cdots\hat{\sigma}(0)\hat{\mu}_{m_{0}}Q\qquad Q(0):=Q
\end{eqnarray}
Then, the function we want to define is:
\begin{eqnarray}\label{zcyl}
Z_{Q,\mathbf{m}}^{\mathrm{cluster}}(\Psi(0),\Psi(t_{f}))&=&\langle \Psi(0)|\hat{\mu}_{m_{0}}\hat{\sigma}(0)\cdots \hat{\mu}_{m_{t_{f}-1}}\hat{\sigma}(t_{f}-1)|\Psi(t_{f})\rangle\nonumber\\
 \langle \Psi(0)|\in \mathcal{H}^{*}_{Q(0)} &,&  | \Psi(t_{f})\rangle\in \mathcal{H}_{Q(t_{f})}
\end{eqnarray}
Now, we return to the problem of describing $\mathcal{H}_{Q}$. Finding a real polarization in terms of the $\mathbf{Y}_{i}$ operators is not easy in general, so we resort on a very practical reparametrization. For this we need to double the amount of operators and introduce the hermitian operators $(\mathbf{u}_{i},\mathbf{p}_{i})$ satisfying the usual Heisenberg algebra with parameter $\hbar/2$:
\begin{eqnarray}
[\mathbf{u}_{i},\mathbf{p}_{j}]=\frac{\hbar}{2} \delta_{ij}
\end{eqnarray}
then, by simply writing
\begin{eqnarray}
\mathbf{Y}_{i}=\mathbf{p}_{i}-Q_{ij}\mathbf{u}_{j}
\end{eqnarray}
the $\mathbf{Y}_{i}$'s defined this way automatically satisfy (\ref{commY}), but we are now working with the variables $(\mathbf{u}_{i},\mathbf{p}_{i})$ which can be identified as usual position and momentum operators. Therefore we can choose a position representation, so $\mathcal{H}_{Q}$ is spanned by the basis $|u\rangle$ or equivalently span $\mathcal{H}_{Q}$ by a momentum basis $|p\rangle$:
\begin{eqnarray}
\langle u |\mathbf{u}_{i}=\langle u|u_{i}\qquad \langle u |\mathbf{p}_{i}=-\frac{\hbar}{2}\frac{\partial}{\partial u_{i}}\langle u|\nonumber\\
\langle p |\mathbf{p}_{i}=\langle p|p_{i}\qquad \langle p |\mathbf{u}_{i}=\frac{\hbar}{2}\frac{\partial}{\partial p_{i}}\langle p|\nonumber\\
\langle u |p \rangle=e^{-\frac{2}{\hbar}u\cdot p}\qquad \mathbf{u}_{i}^{\dag}=\mathbf{u}_{i}\qquad \mathbf{p}_{i}^{\dag}=\mathbf{p}_{i}
\end{eqnarray}
they also should satisfy the completeness relations\footnote{Here the measures are defined as $du=\prod_{i=1}^{N_{0}}du_{i}$ and $dp=\prod_{i=1}^{N_{0}}\frac{dp_{i}}{\hbar}$}:
\begin{eqnarray}
\mathbf{1}=\int_{\mathbb{R}}du| u\rangle\langle u|=\int_{\mathbb{R}}dp| p\rangle\langle p|
\end{eqnarray}
with this Hilbert space at hand we can now write (\ref{zcyl}) in the following way (of course we could have chosen a position basis and the form would have been the same):
\begin{eqnarray}\label{expect}
Z_{Q,\mathbf{m}}^{\mathrm{cluster}}(p(0),p(t_{f}))=\langle p(0)|\hat{\mu}_{m_{0}}\hat{\sigma}(0)\cdots \hat{\mu}_{m_{t_{f}-1}}\hat{\sigma}(t_{f}-1)|p(t_{f})\rangle
\end{eqnarray}
The action of the operators $\hat{\mu}_{k}$ and $\widehat{P}_{k}$ defined before, extends naturally to an action on $\mathbf{p}$ and $\mathbf{u}$. In the following we will only need explicitly the action of $\widehat{P}_{k}$ on $\mathbf{u}$, which is linear \cite{2011SIGMA...7..102K}:
\begin{eqnarray}
\widehat{P}_{k}(\mathbf{u})_{j}= \begin{cases}
    -\mathbf{u}_{k}+\sum_{i}[Q_{i,k}]_{+}\mathbf{u}_{i} & \text{if } j=k\\
     \mathbf{u}_{j}& \text{otherwise }
  \end{cases}
\end{eqnarray}
Using the properties of $\mathbf{p}$ and $\mathbf{u}$ operators, we can bring (\ref{expect}) to the form of an integral (see Appendix B of \cite{Gang:2015wya} for a detailed derivation of this result):
\begin{eqnarray}\label{zL0}
Z_{Q,\mathbf{m}}^{\mathrm{cluster}}(p(0),p(t_{f}))=\int\prod_{t=0}^{t_{f}-1}du(t)\prod_{t=1}^{t_{f}-1}dp(t)\prod_{t=0}^{t_{f}-1}e_{b}\left(\frac{-Y_{m_{t}}(t)}{2\pi b}\right)e^{\frac{2}{\hbar}[u(t)p(t)-\sigma(t)\cdot\widehat{P}_{m_{t}}(u(t))p(t+1)]}
\end{eqnarray}
The total number of integration variables in (\ref{zL0}) is $2t_{f}-1$. In (\ref{zL0}) $u(t)$ and $p(t)$ denote just integration variables. We have left the integration contour undefined, in principle it is the real line, but we may require deforming it for analytic continuation. For the perturbative expansion we will analyze in the following section, we don't need to specify a contour and in general, the question of admissible contours is an open question. The function (\ref{zL0}) can be essentially thought of as the partition function on the mapping cylinder. To get the partition function on the mapping torus, we need to 'wrap it up'. To be more precise, we need to be more specific on the properties of the sequences of mutations and permutations that can be identified with actions of $MCG(\Sigma_{1,1})$. Let's denote $\hat{\varphi}=\mathbf{m}$ the operator corresponding to an element $\varphi\in MCG(\Sigma_{1,1})$. Then the quiver, under the action of $\hat{\varphi}$ returns to the same as the original one: $\hat{\varphi}(Q)=Q$ but the $y_{i}$'s doesn't have to\footnote{The operators $\hat{\varphi}$ acting on the $y_{i}$ variables are required to satisfy the relations of the mapping class group. For example for $\Sigma_{1,1}$, suppose we have a sequence of mutations and permutations $\widehat{S}$ that we associate to $S\in SL(2,\mathbb{Z})$, then $(\widehat{S})^{4}(y)_{i}=y_{i}$.} \cite{FockGoncharovHigher}. Then, under the action of $\hat{\varphi}$, we can identify the Hilbert spaces $\mathcal{H}_{Q}$ and $\mathcal{H}_{Q(t_{f})}$ and so the trace of $\hat{\varphi}$ makes sense. This is the partition function we actually want to associate with the mapping torus:
\begin{eqnarray}
Z_{Q,\hat{\varphi}}^{\mathrm{cluster}}(M_{\varphi})\sim\mathrm{Tr}_{\mathcal{H}_{Q}}\left(\hat{\varphi}\right)
\end{eqnarray}
but we need to make this more precise. By writing $\hat{\varphi}$ as a sequence $\mathbf{m}$ we can construct (\ref{zL0}) associated to $\hat{\varphi}$, then taking the trace seems straightforward: identify $p(0)\simeq p(t_{f})$ and then integrate over $p(0)$. If we do such a thing, we will get a divergent result in general. The reason for this is that the integrand will be independent of some directions in the $p(0)$ variables. This is expected and was already noted for example in \cite{Dimofte:2009yn} and \cite{Dijkgraaf:2010ur} in the context of state integrals models. These directions have actually a very important meaning. At the level of the algebra $A_{Q}$ what is happening is that it's center is not trivial, some combinations of the $\mathbf{Y}_{i}$ variables belong to the center of $A_{Q}$ i.e. they have vanishing commutator with all $\{\mathbf{Y}_{i}\}_{i}^{N_{0}}$. There is a simple way to characterize these elements, denote them by $\mathbf{L}^{\alpha}$:
\begin{eqnarray}
\mathbf{L}^{\alpha}:=c^{\alpha}_{i}\mathbf{Y}_{i}\qquad c^{\alpha}\in \mathrm{Ker}(Q)\qquad \alpha=1,\ldots,n_{C}:=|\mathrm{Ker}(Q)|
\end{eqnarray}
when we are considering $M_{\varphi}$ corresponding to a knot complement, the geometric interpretation of $\mathbf{L}^{\alpha}$ is that they should correspond to the longitudinal holonomy of the knot \cite{Gang:2013sqa,Gang:2015wya} \footnote{Note that, by definition $[\mathbf{L}^{\alpha},\hat{\varphi}]=0$.}. This means that we are working on the polarization where the longitudinal holonomy eigenvalues are taken as position variables. So, our partition function should be interpreted as
\begin{eqnarray}
Z_{Q,\mathbf{m}}^{\mathrm{cluster}}(L^{\alpha})=\langle L^{\alpha}| \Psi \rangle \qquad | \Psi \rangle\in \mathcal{H}_{\partial M_{\varphi}}
\end{eqnarray}
where $\langle L^{\alpha}|\mathbf{L}^{\alpha}=\langle L^{\alpha}|L^{\alpha}$ and $| \Psi \rangle$ is a state determined by $\varphi$. In practice this boils down to taking the trace of (\ref{zL0}) but with the additional insertions of delta functions \footnote{For details of the derivation see Appendix B of \cite{Gang:2015wya}. We remark here that the final result in \cite{Gang:2015wya} looks slightly different because there the number of integration variables have been reduced by using the Fourier transform of the quantum dilogarithm (\ref{FTdilog}). For the purposes of this note, the expression (\ref{pffinal}) will be more suitable.} :
\begin{eqnarray}\label{pffinal}
Z_{Q,\mathbf{m}}^{\mathrm{cluster}}(L^{\alpha})&=&\int \frac{d^{n_{C}}s}{(2\pi i\hbar)^{n_{C}}}\frac{d^{n_{C}}z}{(2\pi i \hbar)^{n_{C}}}\left(\prod_{i}dp_{i}(0)\right)Z_{Q,\mathbf{m}}^{\mathrm{cluster}}(p(0),p(0))\nonumber\\
&&\times e^{-\frac{1}{\hbar}s_{\alpha}(c_{i}^{\alpha}p_{i}(0)-L^{\alpha})-\frac{1}{\hbar}z_{\alpha}c_{i}^{\alpha}u_{i}(0)}
\end{eqnarray}
here, we have written the inserted delta functions $\delta(c_{i}^{\alpha}p_{i}(0)-L^{\alpha})$ and $\delta(c_{i}^{\alpha}u_{i}(0))$ using the formula
\begin{eqnarray}
\int \frac{ds}{2\pi i \hbar}e^{-\frac{1}{\hbar}sx}=\delta(x)
\end{eqnarray}
Before closing this section let's remark on a few important properties of (\ref{pffinal}). One can in principle write $Z_{Q,\mathbf{m}}^{\mathrm{cluster}}$ corresponding to a different choice of polarization of $\mathcal{H}_{\partial M_{\varphi}}$. We will not analyze this problem further here, but for instance, in \cite{Gang:2015wya} there is a more detailed study of the case when the chosen polarization corresponds to use the meridian holonomy eigenvalues $\{M^{\alpha}\}$ as position variables, therefore corresponding to a partition function $Z_{Q,\mathbf{m}}^{\mathrm{cluster}}(M^{\alpha})$. The partition function (\ref{pffinal}), as constructed, using FG coordinates, is expected, and is shown, in various examples in \cite{Gang:2015wya}, to be equivalent to the state integral model partition function studied in \cite{Dimofte:2011gm,Dimofte:2013iv}. However, this is only valid when we construct (\ref{pffinal}) starting from the quivers defined in \cite{FockGoncharovHigher}. The construction presented here carries on for general quivers and sequences of mutations. This fact and some other properties of (\ref{pffinal}) makes it very well suited for applications which are not clear how to implement in the state-integral models of \cite{Dimofte:2011gm,Dimofte:2013iv}. This will be the subject of the next section.

\Section{Applications of cluster partition function}\label{secappl}

In this section we will give an overview of current and potential applications of the cluster partition function (\ref{pffinal}). As a general remark, let's emphasize that (\ref{pffinal}) can in principle be defined for any quiver $Q$ and a given sequence of mutations of $Q$ which leaves the $y_{i}$ variables associated to $Q$ invariant (for defining the expectation value (\ref{expect}) and (\ref{zL0}) this last condition is not even necessary). So, we expect to have a much broader range of applicability than just 3-manifolds $M_{\varphi}$.

\Subsection{Perturbative invariants}\label{pertinvts}

We want to claim that when $M_{\varphi}$ is a hyperbolic knot complement, then we can get perturbative invariants of knots\footnote{However, in general, we expect that we can do the same for any 3-manifold that can be obtained from a mapping torus construction}. In order to study the perturbative invariants defined by (\ref{pffinal}) we proceed as in \cite{Dimofte:2009yn}. First let's determine the critical points that we will use to expand around. Introduce the following shortcut notations:
\begin{eqnarray}
x:=(p,u,s,t,z)\qquad dx:= \frac{d^{n_{C}}s}{(2\pi i\hbar)^{n_{C}}}\prod_{t=0}^{t_{f}-1}du(t)\prod_{t=0}^{t_{f}-1}dp(t)
\end{eqnarray}
note that $x$ includes all variables we are integrating on, this is $p_{i}(t),u_{j}(t),s_{\alpha},z_{\alpha}$ for all $t=0,\ldots,t_{f}-1$, $i,j=1,\ldots,N_{0}$ and $\alpha=1,\ldots, n_{C}$. At leading order when $\hbar\rightarrow 0$ we expect that $Z_{Q,\mathbf{m}}^{\mathrm{cluster}}(L^{\alpha})$ takes the form
\begin{eqnarray}
Z_{Q,\mathbf{m}}^{\mathrm{cluster}}(L^{\alpha})\sim\int dx e^{\frac{1}{\hbar}V(x;L)}
\end{eqnarray}
with use of (\ref{ebexpansion}) this can be calculated explicitly and it yields:
\begin{eqnarray}\label{ourpotential}
V(x;L)=-z_{\alpha}c^{\alpha}(u(0))-s_{\alpha}(c^{\alpha}(p(0))-L^{\alpha})+2\sum_{t=0}^{t_{f}-1}[u(t)p(t)-\sigma(t)\cdot\widehat{P}_{m_{t}}(u(t))p(t+1)]+\sum_{t=0}^{t_{f}-1}\mathrm{Li}_{2}(-e^{-Y_{m_{t}}(t)})
\end{eqnarray}
where $p(L):=p(0)$ and recall that:
\begin{eqnarray}
Y_{m_{t}}(t)=p_{m_{t}}(t)-Q_{m_{t},j}(t)u_{j}(t)
\end{eqnarray}
so, the critical points of $V(x;L)$ (i.e. $x$ such that $\partial_{x}V=0$) will be determined by the following equations:
\begin{eqnarray}\label{criteqs}
\frac{\partial V}{\partial s_{\alpha}}&=&-(c^{\alpha}(p(0))-L^{\alpha})\qquad\frac{\partial V}{\partial z_{\alpha}}=-c^{\alpha}(u(0))\nonumber\\
\frac{\partial V}{\partial p_{i}(t)}&=&-\delta_{t,0}s_{\alpha}c^{\alpha}_{i}+2\left(u_{i}(t)-(\sigma(t-1)\cdot \widehat{P}_{m_{t-1}}(u(t-1)))_{i}\right)+\delta_{m_{t},i}\ln\left(1+e^{-Y_{m_{t}}(t)}\right)\nonumber\\
\frac{\partial V}{\partial u_{i}(t)}&=&-\delta_{t,0}z_{\alpha}c^{\alpha}_{i}+2\left(p_{i}(t)-(\sigma(t)\cdot \widehat{P}_{m_{t}})_{ji}p_{j}(t+1)\right)-Q_{m_{t},i}(t)\ln\left(1+e^{-Y_{m_{t}}(t)}\right)
\end{eqnarray}
where $\sigma(-1)\cdot \widehat{P}_{m_{-1}}(u(-1)):=\sigma(t_{f}-1)\cdot \widehat{P}_{m_{t_{f}-1}}(u(t_{f}-1))$. So, fix a critical point $x^{(c)}\in \mathrm{Crit}(V)$ i.e. a solution of the equations (\ref{criteqs}), then, we can perform a Feynman diagram expansion around $x^{(c)}$ \footnote{For the readers not familiar with this, we recommend the review \cite{2004math......6251P}.}. Before we proceed it is convenient to make a slight change of variables in $Z_{Q,\mathbf{m}}^{\mathrm{cluster}}(L^{\alpha})$, we shift the $p_{m_{t}}(t)$ variables by $Q_{m_{t},j}(t)u_{j}(t)$ for $t=0,\ldots, t_{f}-1$ and then rename to change
\begin{eqnarray}
Y_{m_{t}}(t)\rightarrow p_{m_{t}}(t)
\end{eqnarray}
since this change of variables plus renaming doesn't introduce any extra factors (the determinant of the Jacobian of this transformation is $1$), we obtain:
\begin{eqnarray}
Z_{Q,\mathbf{m}}^{\mathrm{cluster}}(L^{\alpha})&=&\int dx e^{-\frac{1}{\hbar}s_{\alpha}(c_{i}^{\alpha}p_{i}(0)-L^{\alpha})-\frac{1}{\hbar}z_{\alpha}c_{i}^{\alpha}u_{i}(0)}\prod_{t=0}^{t_{f}-1}e_{b}\left(\frac{-p_{m_{t}}(t)}{2\pi b}\right)e^{\frac{2}{\hbar}[u(t)p(t)-\sigma(t)\cdot\widehat{P}_{m_{t}}(u(t))p(t+1)+F_{m_{t}}]}\nonumber\\
F_{m_{t}}&:=& u_{m_{t}}(t)Q_{m_{t},j}(t)u_{j}(t)-\left(\sigma(t)\cdot\widehat{P}_{m_{t}}(u(t))\right)_{m_{t+1}}Q_{m_{t+1},j}(t+1)u_{j}(t)-\frac{\delta_{t,0}}{2}c_{m_{t}}^{\alpha}Q_{m_{t},j}(t)u_{j}(t)\nonumber\\ t_{f}&\simeq& 0
\end{eqnarray}
note that we can more compactly write:
\begin{eqnarray}\label{newZ}
Z_{Q,\mathbf{m}}^{\mathrm{cluster}}(L^{\alpha})=\int dx e^{\frac{1}{\hbar}V(x;L)}\prod_{t=0}^{t_{f}-1}e_{b}\left(\frac{-p_{m_{t}}(t)}{2\pi b}\right)e^{-\frac{1}{\hbar}\mathrm{Li}_{2}\left(-e^{-p_{m_{t}}(t)}\right)}\nonumber\\
\end{eqnarray}
where $V(x;L)$ in (\ref{newZ}) is understood as the potential after the change of variables we just defined. Now we can proceed with the perturbative expansion. For this we write $x=x^{(c)}+\tilde{x}$, where $\tilde{x}$ is a small perturbation and then use (\ref{ebexpansion}) to write $Z_{Q,\mathbf{m}}^{\mathrm{cluster}}(L^{\alpha})$ as
\begin{eqnarray}\label{Zexp}
Z_{Q,\mathbf{m}}^{\mathrm{cluster}}(L^{\alpha})^{(c)}=e^{\Gamma^{(0)}}\int d\tilde{x} e^{\frac{1}{2\hbar}H_{ab}\tilde{x}^{a}\tilde{x}^{b}}\prod_{t=0}^{t_{f}-1}e^{\sum_{k\geq 1}\frac{(-1)^{k}}{k!}\Gamma_{t}^{(k)}\left(\tilde{p}_{m_{t}}(t)\right)^{k}}
\end{eqnarray}
where
\begin{eqnarray}\label{Fdiags}
H_{ab}&=&\frac{\partial^{2}V(x;L)}{\partial x^{a}\partial x^{b}}\Big|_{x=x^{(c)}}\nonumber\\
\Gamma^{(0)}&=&\frac{1}{\hbar}V(x^{(c)};L)+\sum_{n\geq 1}\left(\frac{B_{n}}{2^{n-1}}-1\right)\frac{\hbar^{n-1}}{n!}\mathrm{Li}_{2-n}\left(-e^{-p^{(c)}_{m_{t}}(t)}\right)\nonumber\\
\Gamma^{(k)}_{t}&=&\sum_{n\geq n_{k}}\left(\frac{B_{n}}{2^{n-1}}-1\right)\frac{\hbar^{n-1}}{n!}\mathrm{Li}_{2-n-k}\left(-e^{-p^{(c)}_{m_{t}}(t)}\right)
\end{eqnarray}
and
\begin{eqnarray}
n_{k}=\begin{cases}
    1& k=1,2\\
    0 & k>2 \\
  \end{cases}
\end{eqnarray}
Whit this information, from (\ref{Zexp}) we can read the perturbative expansion of $Z_{Q,\mathbf{m}}^{\mathrm{cluster}}(L^{\alpha})^{(c)}$ and indeed, one can see that it takes the form (\ref{pertexp}) where the index $\alpha$ in (\ref{pertexp}) is labelling the choice of critical point $x^{(c)}$. The situation is very analogous to \cite{Dimofte:2012qj}, so we can use their results\footnote{We remark here that the integration in $d\tilde{x}$ in $Z_{Q,\mathbf{m}}^{\mathrm{cluster}}(L^{\alpha})^{(c)}$ is Gaussian in all variables but $\{\tilde{p}_{m_{t}}(t)\}_{t=0}^{t_{f}-1}$. So, higher order invariants will depend only on these latter variables.}. Let's first read the lower $\hbar$ degree terms in (\ref{pertexp}). Given a Laurent series $f(\hbar)$ on $\hbar$, denote $\mathrm{\mathrm{coeff}}[f(\hbar),\hbar^{a}]$ the coefficient of $\hbar^{a}$ in $f(\hbar)$. Then, first terms in (\ref{Zexp}) read:
\begin{eqnarray}
S^{(c)}_{0}&=&\mathrm{coeff}[\Gamma^{(0)},\hbar^{-1}]\nonumber\\
\exp(S_{1}^{(c)})&=&\frac{(-1)^{t_{f}N_{0}}(2\pi)^{t_{f}N_{0}}-n_{C}}{\sqrt{\mathrm{det}H}}e^{\mathrm{coeff}[\Gamma^{(0)},\hbar^{0}]}
\end{eqnarray}
the $\ln\hbar$ term in (\ref{pertexp}) and (\ref{Zexp}), comes from an overall normalization that we should fix. Actually the only terms affected by the overall normalization of (\ref{Zexp}) are the $\hbar^{0}$ and $\ln\hbar$ terms. We will ignore this issue here and move on to the $S_{n}$ terms. They are computed by a Feynman diagram expansion. From (\ref{Fdiags}) we can immediately see that we will have vertices of all valences $k=1,2,3,\ldots$ Use the indices $t$ to denote the indices corresponding to the coordinates $\{p_{m_{t}}(t)\}_{t=0}^{t_{f}-1}$. Then, define the propagator:
\begin{eqnarray}
\Pi_{t,t'}:=-\hbar(H^{-1})_{t,t'}
\end{eqnarray}
The terms $S_{n}$ will be extracted from a sum of connected graphs. Consider $\mathcal{G}_{\Gamma}$ a connected graph with vertices of valences $k\geq 1$. Then, we associate a weight to $\mathcal{G}_{\Gamma}$: to each $k$-vertex we associate a factor $\Gamma^{(k)}_{t_{i}}$ and a label $t_{i}$ and to each internal line connecting two vertices with labels $t_{i}$ and $t_{j}$ a factor $\Pi_{t_{i},t_{j}}$. Then we can define the weight associated to $\mathcal{G}_{\Gamma}$:
\begin{eqnarray}
W_{\Gamma}(\mathcal{G}_{\Gamma}):=\frac{1}{|\mathrm{Aut}(\mathcal{G}_{\Gamma})|}\sum_{\mathrm{labels}}\prod_{v\in \mathrm{vertices}}(-1)^{k_{v}}\Gamma^{(k_{v})}_{t_{v}}\prod_{e\in \mathrm{edges}}\Pi_{e}
\end{eqnarray}
Where $|\mathrm{Aut}(\mathcal{G}_{\Gamma})|$ is the symmetry factor (the rank of the group of automorphisms of $\mathcal{G}_{\Gamma}$). Given a connected graph $\mathcal{G}_{\Gamma}$ then is easy to see that $W_{\Gamma}(\mathcal{G}_{\Gamma})$ is of order $\hbar^{-V+E}$ or higher, where $E$ is the number of internal lines and $V$ the number of vertices with valence $k\geq 3$ in $\mathcal{G}_{\Gamma}$. After some computation one can show that $E=V+\mathcal{L}+V_{1}+V_{2}-1$ where $\mathcal{L}$ is the number of loops and $V_{1}$, $V_{2}$ are the number of $1$ and $2$-vertices respectively. Then, if we define:
\begin{eqnarray}
\mathcal{G}_{n}:=\{ \text{  Connected graphs \ }\mathcal{G}_{\Gamma}\text{ \ such that \ }\mathcal{L}+V_{1}+V_{2}\leq n \}
\end{eqnarray}
then
\begin{eqnarray}
S^{(c)}_{n}=\mathrm{coeff}\left[\Gamma^{(0)}+\sum_{\mathcal{G}_{\Gamma}\in \mathcal{G}_{n}}W_{\Gamma}(\mathcal{G}_{\Gamma}),\hbar^{n-1}\right]\qquad n\geq 2
\end{eqnarray}
This concludes our analysis of perturbative invariants defined from $Z_{Q,\mathbf{m}}^{\mathrm{cluster}}(L^{\alpha})$. We point out that in \cite{2013arXiv1310.3068K}, $S^{(c)}_{1}$ has been associated with topological invariants of knots. We expect $S^{(c)}_{n\geq 2}$ to provide novel invariants of knots or cluster variables in general.

\Subsection{Exploring other branches of $\mathcal{M}_{\mathrm{flat}}$}\label{missingbr}

As previously mentioned, the partition function $Z_{Q,\mathbf{m}}^{\mathrm{cluster}}(L^{\alpha})$ constructed using the quiver and coordinates specified in \cite{FockGoncharovHigher} is expected to correspond to a subspace of $\bigcup_{\rho}\mathcal{M}_{\mathrm{flat}}(\Sigma_{1,1},\rho)$. This is reflected in the fact that the partition function $Z_{Q,\mathbf{m}}^{\mathrm{cluster}}(L^{\alpha})$ is not the most general solution to the equations (\ref{ApolyN}) when $\widehat{A}_{i}(\hat{l},\hat{m})$ are operators obtained by quantization of the polynomial equations defining the classical solutions to the CS equations of motion. For the case of knot complements and $N=2$ there is only one equation $A(l,m)=0$ given by the A-polynomial of the knot \cite{CooperApolynomial}. All the A-polynomials have a factor $(l-1)$ where $l$ is the eigenvalue of the longitudinal holonomy. Even though it is not yet proven rigourously, there is overwhelming evidence that for these cases the state-integral model of \cite{Dimofte:2011gm} (and, hence the cluster partition function with FG quiver), as furthermore analyzed in \cite{Dimofte:2012qj} is only missing the branch given by $l=1$, also known as the 'abelian branch', i.e. all these partition functions are annihilated by a quantization of $A(l,m)/(l-1)$. From a physics perspective, it is expected that one can construct a partition function which gets contributions from all branches \cite{Lee:2013ida,Yagi:2013fda,Cordova:2013cea,Dimofte:2014ija,Chung:2014qpa}. A proposal for a theory, in the case of $N=2$, which can contain all the possible branches is analyzed in \cite{Chung:2014qpa}.\\

However, even at the classical level, analyzing the space of flat $SL(N,\mathbb{C})$-connections for $N>2$ is a considerably more difficult task and have been carried out only in some specific cases. See for instance \cite{2014arXiv1412.4711F,2015arXiv150504451H}. Therefore, a quantization is not, in general, available.\\

For $N>2$ and higher, the problem of missing branches is even more severe. The abelian branch is still missing but also there are new missing branches and they grow with $N$ (see for example section 7.3 in \cite{Dimofte:2013iv}). There is no systematic analysis on a precise characterization of these missing branches, but physics give us a guide: for knot complements, the longitudinal holonomies $\rho$ are labelled by an embedding of $SU(2)\rightarrow SU(N)$ \cite{Gaiotto:2008ak}, this corresponds to partitions of $N$ of the form (we pick an ordering $n_{a}\geq n_{a+1}$):
\begin{eqnarray}
\rho=[n_{1},\ldots,n_{s}]\qquad \sum_{a=1}^{s}n_{a}=N\qquad n_{a}\in \mathbb{Z}_{\geq 1}
\end{eqnarray}
note that we are using $\rho$ to denote a holonomy as well as a partition of $N$. In other words we are fixing $\rho$ in $\mathcal{M}_{\mathrm{flat}}(\Sigma_{1,1},\rho)$. These different choices of $\rho$ are known in the physics literature as 'codimension $2$ defects'. For the FG quiver case, this corresponds to the choice $\rho=[1,\ldots,1]$. For $N=2$, the only option is $\rho=[1,1]$, but for $N>2$ more options arise and at the time this note is written, there is no known systematic way to compute partition functions of these other cases in general. One of the main claims in \cite{Gang:2015wya} is that the branches corresponding to different choices of partitions of $N$ can be implemented by changing the quiver $Q$ in $Z_{Q,\mathbf{m}}^{\mathrm{cluster}}$. In \cite{Gang:2015wya}, some quivers are worked out for some families of $\rho$'s, taking as inspiration the work of \cite{Xie:2012dw}. Some very nontrivial consistency checks are preformed but it is still an open question how to determine the quiver corresponding to a given $\rho$ in a systematic way. Given $\rho$ there are some immediate conditions one can ask for a quiver $Q_{\rho}$ to be associated to it, these are
\begin{eqnarray}
|Q_{\rho}|&=&\mathrm{dim}_{\mathbb{C}}\mathcal{M}_{\mathrm{flat}}(\Sigma_{1,1},\rho)+|\mathrm{Ker}(Q_{\rho})|\nonumber\\
|\mathrm{Ker}(Q_{\rho})|&=&\sum_{\alpha}l_{\alpha}-1
\end{eqnarray}
where $l_{\alpha=1,2,\ldots}$ denotes the number of times $\alpha$ appears in $\rho=[n_{1},\ldots,n_{s}]$, $|Q_{\rho}|$ is the number of vertices of the quiver and
\begin{eqnarray}
\mathrm{dim}_{\mathbb{C}}\mathcal{M}_{\mathrm{flat}}(\Sigma_{1,1},\rho)=(N^{2}-2)-\sum_{i=1}^{s}(n_{i}^{2}-1)
\end{eqnarray}
Therefore we expect that for a quiver $Q_{\rho}$, $Z_{Q_{\rho},\mathbf{m}}^{\mathrm{cluster}}(L^{\alpha})$ depends on parameters $L^{\alpha}$ with $\alpha=1,\ldots,|\mathrm{Ker}(Q_{\rho})|$. Formulas for more general Riemann surfaces $\Sigma$ are given in \cite{Gang:2015wya}. We should remark that by this method, it is still unclear how to obtain a partition function capturing the abelian branch. Another open problem is, given a proposed quiver $Q_{\rho}$ with $\rho\neq [1,\ldots,1]$, how to find sequences of mutations implementing the action of $MCG(\Sigma)$ on the corresponding $y_{i}$ coordinates. This is done in some examples in \cite{Gang:2015wya}, but a general, systematic method is lacking.

\Subsection{Wilson loop insertions}

The insertion of Wilson loops along knots for CS theories with compact gauge group $H$, reviewed briefly in section 2, is a well established and active subject of research in physics and mathematics, starting with the pioneering work \cite{Witten:1988hf} where expectation values of Wilson loops along knots $\mathcal{K}$ embedded in $S^{3}$ were related to the (colored) Jones polynomial of $\mathcal{K}$ . A Wilson loops is determined by the isotopy class of the knot $\mathcal{K}$ and an unitary representation $R$ of $H$. More precisely, the Wilson loop operator is given by the trace on $R$ of the holonomy of the connection along $\mathcal{K}$:
\begin{eqnarray}
W_{R}(\mathcal{K}):=\mathrm{Tr}_{R}Pe^{\oint_{\mathcal{K}}A}
\end{eqnarray}
where $P$ stands for the path ordering operator. The expectation value of $W_{R}(\mathcal{K})$ is then computed by \footnote{We will work here with un-normalized expectation values i.e., we will not divide them by the partition function.}
\begin{eqnarray}
\langle W_{R}(\mathcal{K})\rangle_{S^{3}}=\int_{\mathcal{Y}_{H}}\mathcal{D}Ae^{\frac{i}{4\pi}kCS[A]}W_{R}(\mathcal{K})
\end{eqnarray}
where
\begin{eqnarray}
CS[A]:=\int_{S^{3}}\mathrm{Tr}\left(AdA+\frac{2}{3}A\wedge A\wedge A\right)
\end{eqnarray}
Alternatively one can consider CS theory on the knot complement $S^{3}\setminus \mathcal{K}$ and fix boundary conditions in terms of the holonomy of $A$ along a non-trivial cycle of the boundary torus. There is a one-to-one correspondence between boundary conditions and expectation values of Wilson loops for the case of compact gauge group (see for example \cite{Elitzur:1989nr}).\\

In contrast, for the case of complex gauge group $G$, the correspondence between boundary conditions and Wilson loop expectation values is not completely well established. For instance, when $G$ is noncompact, infinite dimensional representations exist that don't come from a lift of a unitary representation of a real form of $G$. For a discussion about this in the case of $G=SL(2,\mathbb{C})$ see \cite{Gukov:2003na} and \cite{2010arXiv1003.4808D}. Motivated by physics considerations we would like to consider the problem of computing CS partition function, for gauge group $G$, on a knot complement $M=S^{3}\setminus \mathcal{K}$ in the presence of a Wilson loop along a nontrivial cycle $\gamma\in\pi_{1}(S^{3}\setminus \mathcal{K})$ \cite{Gang:2015wya}. Formally, this corresponds to computing
\begin{eqnarray}
\langle W_{R}(\gamma)\rangle_{M}=\int_{\mathcal{Y}}\mathcal{DA}\mathcal{D\overline{A}}e^{iS_{CS}}\mathrm{Tr}_{R}P e^{\oint_{\gamma}\mathcal{A}}
\end{eqnarray}
where $R$ is the holomorphic lift of a finite dimensional representation of $H\subset G$, where $H$ is a maximal compact subgroup of $G$. In the cases we consider, our knot complement comes from a mapping torus construction: $M=(\Sigma_{1,1}\times S^{1})_{\varphi}$. Take $\gamma\in \pi_{1}(\Sigma_{1,1})$ and consider its natural lift to $\pi_{1}(M)$. When working in the FG coordinates associated to $G=SL(N,\mathbb{C})$ and $\Sigma_{1,1}$, there is a prescription to compute the holonomy $\mathrm{Hol}_{\gamma}(\mathcal{A})$, up to a $SL(N,\mathbb{C})$ gauge transformation which acts by conjugation on $\mathrm{Hol}_{\gamma}(\mathcal{A})$ \cite{FockGoncharovHigher}. Therefore, we can compute the classical value of $\mathrm{Tr}_{R}(\mathrm{Hol}_{\gamma}(\mathcal{A}))$ in terms of the $y_{i}$ coordinates. It takes the following form:
\begin{eqnarray}
W_{R}(\gamma):=\mathrm{Tr}_{R}(\mathrm{Hol}_{\gamma}(\mathcal{A}))=\sum_{k}c_{k}e^{\sum_{i=1}^{N_{0}}a_{i}Y_{i}} \qquad c_{k},a_{i}\in\mathbb{Z}
\end{eqnarray}
The function $W_{R}(\gamma)$ can, in principle, be promoted to an operator in terms of $\mathbf{y}_{i}$ and the quantization parameter $q$. For this class of Wilson loops this problem was addressed in \cite{Coman:2015lna} and the answer takes the form:
\begin{eqnarray}
\widehat{W}_{R}(\gamma;\mathbf{Y}_{i})=\sum_{k}\hat{c}_{k}(q)e^{\sum_{i=1}^{N_{0}}a_{i}\mathbf{Y}_{i}}
\end{eqnarray}
where $\hat{c}_{k}(q)$ are (commutative) functions of $q$ such that $\lim_{q\rightarrow 1}\hat{c}_{k}(q)=c_{k}$. Determining $\hat{c}_{k}(q)$ is a not an easy task but we will not need it here. In \cite{Gang:2015wya} we propose the following prescription for computing the expectation value of $W_{R}(\gamma)$:
\begin{eqnarray}
\langle W_{R}(\gamma)\rangle_{M}&=&\int\frac{d^{n_{C}}s}{(2\pi i\hbar)^{n_{C}}}\frac{d^{n_{C}}z}{(2\pi i \hbar)^{n_{C}}}\left(\prod_{i}dp_{i}(0)\right)\langle p(0)|\widehat{W}_{R}(\gamma;\mathbf{Y}_{i}(0))\hat{\mu}_{m_{0}}\hat{\sigma}(0)\cdots \hat{\mu}_{m_{t_{f}-1}}\hat{\sigma}(t_{f}-1)|p(0)\rangle \nonumber\\
&\times& e^{-\frac{1}{\hbar}s_{\alpha}(c_{i}^{\alpha}p_{i}(0)-L^{\alpha})-\frac{1}{\hbar}z_{\alpha}c_{i}^{\alpha}u_{i}(0)}
\end{eqnarray}
this can be reduced using the same techniques that we applied to the case without the insertion of $W_{R}(\gamma)$. Here we present the case of just one exponential $\exp(\sum_{i=1}^{N_{0}}a_{i}\mathbf{Y}_{i})$, and $\langle W_{R}(\gamma)\rangle_{M}$ is easily obtained by linear extension:
\begin{eqnarray}\label{wilsonloopa}
\langle e^{\sum_{i=1}^{N_{0}}a_{i}Y_{i}} \rangle_{M}&=&\int dx e^{-\frac{1}{\hbar}s_{\alpha}(c_{i}^{\alpha}p_{i}(0)-L^{\alpha})-\frac{1}{\hbar}z_{\alpha}c_{i}^{\alpha}u_{i}(0)+a\cdot p(0)}\prod_{t=0}^{t_{f}-1}e_{b}\left(\frac{-p_{m_{t}}(t)}{2\pi b}\right)e^{\frac{2}{\hbar}[u(t)p(t)-\sigma(t)\cdot\widehat{P}_{m_{t}}(u(t))p(t+1)+F_{m_{t}}(a)]}\nonumber\\
F_{m_{t}}(a)&:=& u_{m_{t}}(t)Q_{m_{t},j}(t)u_{j}(t)-\left(\sigma(t)\cdot\widehat{P}_{m_{t}}(u(t))\right)_{m_{t+1}}Q_{m_{t+1},j}(t+1)u_{j}(t)
-\frac{\delta_{t,0}}{2}c_{m_{t}}^{\alpha}Q_{m_{t},j}(t)u_{j}(t)\nonumber\\ &+&\delta_{t,t_{f}}\frac{\hbar}{2}\left(\sigma(t-1)\cdot\widehat{P}_{m_{t-1}}(u(t-1))\right)_{i}Q_{i,k}(0)a_{k}\nonumber\\ t_{f}&\simeq& 0
\end{eqnarray}
for the details on the derivation of this result we refer the reader to \cite{Gang:2015wya}. Notice that we can carry on a perturbative analysis in a similar fashion as in the previous section. First note that (\ref{wilsonloopa}) can be written as
\begin{eqnarray}
\langle e^{\sum_{i=1}^{N_{0}}a_{i}Y_{i}} \rangle_{M}=\int dx e^{-\frac{1}{\hbar}V(x;L)+x^{b}\mathcal{V}_{b}}\prod_{t=0}^{t_{f}-1}e_{b}\left(\frac{-p_{m_{t}}(t)}{2\pi b}\right)e^{-\frac{1}{\hbar}\mathrm{Li}_{2}\left(-e^{-p_{m_{t}}(t)}\right)}
\end{eqnarray}
this is exactly the same as eq. (\ref{newZ}) with the addition of the term $x^{b}\mathcal{V}_{b}$ which is defined by:
\begin{eqnarray}
x^{b}\mathcal{V}_{b}=a_{i}p_{i}(0)+\left(\sigma(t_{f}-1)\cdot\widehat{P}_{m_{t_{f}-1}}(u(t_{f}-1))\right)_{i}Q_{i,k}(0)a_{k}
\end{eqnarray}
Therefore, the critical points $x^{(c)}$ are the same as in the previous analysis without the insertion of the Wilson loop. Then, we can write the partition function around $x^{(c)}$, as in (\ref{Zexp}):
\begin{eqnarray}\label{intWL}
\langle e^{\sum_{i=1}^{N_{0}}a_{i}Y_{i}} \rangle^{(c)}_{M}=e^{\Gamma^{(0)}(a)}\int d\tilde{x} e^{\frac{1}{2\hbar}H_{ab}\tilde{x}^{a}\tilde{x}^{b}+\mathcal{V}_{a}\tilde{x}^{a}}\prod_{t=0}^{t_{f}-1}e^{\sum_{k\geq 1}\frac{(-1)^{k}}{k!}\Gamma_{t}^{(k)}\left(\tilde{p}_{m_{t}}(t)\right)^{k}}
\end{eqnarray}
where vertices $\Gamma_{t}^{(k)}$, and propagator $H_{ab}$ are the same as in (\ref{Fdiags}) and the modified vacuum energy is
\begin{eqnarray}
\Gamma^{(0)}(a)=\Gamma^{(0)}+\mathcal{V}^{T}x^{(c)}
\end{eqnarray}
In order to write the modified invariants, which we denote $\{S^{(c)}_{n}(a)\}_{n\geq 0}$, it is convenient to perform the shift:
\begin{eqnarray}
\tilde{x}\rightarrow \tilde{x}-\hbar H^{-1}\mathcal{V}
\end{eqnarray}
in particular, we denote the shift of $\tilde{p}_{m_{t}}(t)$ by:
\begin{eqnarray}
\tilde{p}_{m_{t}}(t)\rightarrow \tilde{p}_{m_{t}}(t)-\hbar V_{t}\qquad V_{t}:=\sum_{a}(H^{-1})_{m_{t},a}\mathcal{V}_{a}
\end{eqnarray}
this shift, will get rid of the term $\mathcal{V}_{a}\tilde{x}^{a}$ in the integral (\ref{intWL}), hence all vertices will depend only on $\{\tilde{p}_{m_{t}}(t)\}_{t=0}^{t_{f}-1}$ and the integration will be Gaussian on all the other variables. Therefore, the perturbative analysis can be carried on in a completely analogous way as before, but with modified vertices:
\begin{eqnarray}
\langle e^{\sum_{i=1}^{N_{0}}a_{i}Y_{i}} \rangle^{(c)}_{M}=e^{\Gamma^{(0)}(a)+\frac{\hbar}{2}\mathcal{V}^{T}H^{-1}\mathcal{V}}\int d\tilde{x} e^{\frac{1}{2\hbar}H_{ab}\tilde{x}^{a}\tilde{x}^{b}}\prod_{t=0}^{t_{f}-1}e^{\sum_{k\geq 1}\frac{(-1)^{k}}{k!}\widetilde{\Gamma}_{t}^{(k)}\left(\tilde{p}_{m_{t}}(t)\right)^{k}}
\end{eqnarray}
\begin{eqnarray}
\widetilde{\Gamma}_{t}^{(k)}=\sum_{s\geq k}\frac{(V_{t})^{s-k}}{(s-k)!}\sum_{n\geq n_{s}}\left(\frac{B_{n}}{2^{n-1}}-1\right)\frac{\hbar^{n-1+s-k}}{n!}\mathrm{Li}_{2-n-s}\left(-e^{-p^{(c)}_{m_{t}}(t)}\right)
\end{eqnarray}
Note that the number of terms in $\widetilde{\Gamma}_{t}^{(k)}$ for a fixed degree in $\hbar$, is finite.\\

A couple of remarks are in order. Consistency of this result in some nontrivial cases was analyzed in \cite{Gang:2015wya} and it was found it agrees with physics predictions. There are still various natural generalizations of this result that one can try to address. As we mentioned, the construction of $W_{R}(\gamma)$ relied on the fact that there exists an explicit expression in terms of the FG coordinates, however, for more general quivers associated with other regions of $\bigcup_{\rho}\mathcal{M}_{\mathrm{flat}}(\Sigma_{1,1},\rho)$ such a result is not available. Evidently, for a general quiver, one can write an operator of the form $\sum_{k}c_{k}\exp(a^{(k)}\cdot\mathbf{Y})$ in terms of the corresponding cluster coordinates but its interpretation is not clear. Another possible generalization is changing $\gamma$ by a more general cycle in $\pi_{1}(M)$. Explicit expressions for holonomies of cycles in $\pi_{1}(M)$ exist \cite{Dimofte:2013iv} but their quantization is not known, so, is not clear how to promote them to operators. Ultimately, the goal will be to take $\gamma$ to be another knot, linked to $\mathcal{K}$, but it seems there are many more fundamental problems to solve before we can take that step.

\bigskip

{\bf Acknowledgement}\ \ We would like to thank D. Gang, N. Kim and M. Yamazaki for collaboration in the
works \cite{Gang:2015wya,Gang:2015bwa} on which this note is based. We also thank M. Gabella and S. Lee for useful discussions. We
acknowledge the support from the Institute for Advanced Study and from DOE grant DE-SC0009988.

\appendix

\Section{Definition and some properties of quantum dilogarithm function}

The quantum dilogarithm function $e_{b}(z)$ was originally defined in \cite{FaddeevKashaevQuantum,Faddeev95} and for further properties as well as connection with other special functions we refer to some papers relevant in our context: \cite{FaddeevModular,Faddeev:2000if,Dimofte:2009yn}. By defining the symbol
\begin{eqnarray}
(x;q)_{\infty}=\prod_{k=0}^{\infty}(1-xq^{k})
\end{eqnarray}
we can define $e_{b}(z)$ as quotient of infinite products:
\begin{eqnarray}
e_{b}(z)=\begin{cases}
    \frac{(e^{2\pi b(z+i\mathcal{Q}/2)};q)_{\infty}}{(e^{2\pi b^{-1}(z-i\mathcal{Q}/2)};\tilde{q}^{-1})_{\infty}} &\Im(b^{2})>0\\
    & \\
    \frac{(e^{2\pi b^{-1}(z+i\mathcal{Q}/2)};\tilde{q})_{\infty}}{(e^{2\pi b(z-i\mathcal{Q}/2)};q^{-1})_{\infty}}&\Im(b^{2})<0\\
  \end{cases}
\end{eqnarray}
where
\begin{eqnarray}
 q=e^{2\pi i b^{2}}\qquad\tilde{q}=e^{2\pi ib^{-2}}\qquad \mathcal{Q}=b+b^{-1}
\end{eqnarray}
$e_{b}(z)$ is a meromorphic function for all values of $b$ such that $b^{2}\not\in \mathbb{R}_{\leq 0}$ with zeroes and poles at
\begin{eqnarray}
\text{ \ poles: \ }i\frac{\mathcal{Q}}{2}+i\mathbb{N}b+i\mathbb{N}b^{-1}\qquad\text{ \ zeros: \ }-i\frac{\mathcal{Q}}{2}-i\mathbb{N}b-i\mathbb{N}b^{-1}
\end{eqnarray}
For $|\Im z|<|\Re \mathcal{Q}/2|$, we have an integral expression:
\begin{eqnarray}
e_{b}(z)=\exp\left[\int_{\mathbb{R}+i0}\frac{dt}{4t}\frac{e^{-2itz}}{\sinh(bt)\sinh(b^{-1}t)}\right]=e^{\frac{i\pi}{2}z^{2}+\frac{i\pi}{24}(b^{2}+b^{-2})}s_{b}(z)
\end{eqnarray}
and its most relevant properties for us are, the periodicity:
\begin{eqnarray}
e_{b}(z\pm ib)=(1+e^{2\pi b z}(e^{i\pi b^{2}})^{\pm 1})^{\mp 1}e_{b}(z)\qquad e_{b}(z\pm ib^{-1})=(1+e^{2\pi b^{-1} z}(e^{i\pi b^{-2}})^{\pm 1})^{\mp 1}e_{b}(z)
\end{eqnarray}
and its Fourier transform:
\begin{eqnarray}\label{FTdilog}
\int_{\mathbb{R}}dx e_{b}(x)e^{2\pi i wx}=e^{-i\pi w^{2}+i\pi (1-4c_{b}^{2})/12}e_{b}(w+c_{b})\qquad c_{b}=i\frac{\mathcal{Q}}{2}
\end{eqnarray}
Finally, let us remark on the asymptotic expansion of $e_{b}(z)$, as $b\sim 0 (\Leftrightarrow \hbar=2\pi i b^{2}\sim 0)$ (see \cite{Dimofte:2009yn})
\begin{eqnarray}\label{ebexpansion}
e_{b}(-Y/(2\pi b))\sim\exp\left(\sum_{n=0}^{\infty}\frac{B_{n}}{2^{n-1}n!}\hbar^{n-1}\mathrm{Li}_{2-n}(-e^{-Y})-\sum_{n=0}^{\infty}\frac{\hbar^{n-1}}{n!}\mathrm{Li}_{2-n}(-e^{-Y})\right)
\end{eqnarray}
here $B_{n}$ are the Bernoulli numbers and we are using the convention $B_{1}=\frac{1}{2}$. The polylogarithm function $\mathrm{Li}_{s}(x)$ is defined by
\begin{eqnarray}
\mathrm{Li}_{s}(x)=\sum_{k=1}^{\infty}\frac{x^{k}}{k^{s}} \qquad s\in \mathbb{N}
\end{eqnarray}
the function $\mathrm{Li}_{s}(x)$ for $s<0$ is given by
\begin{eqnarray}
\mathrm{Li}_{-s}(x)=\left( x\frac{\partial}{\partial x} \right)^{s} \frac{x}{1-x}=\sum_{k=0}^{s} k!S(s+1,k+1) \left(\frac{x}{1-x} \right)^{k+1}\in \mathbb{N}\left[\frac{x}{1-x}\right]
\end{eqnarray}
where $S(n,k)$ are Stirling numbers of the second kind.















\bibliographystyle{fullsort}
\bibliography{procbib}

\def\cprime{$'$}
\providecommand{\href}[2]{#2}\begingroup\raggedright\begin{thebibliography}{10}

\bibitem{Dimofte:2014ija}
T.~Dimofte, ``{3d Superconformal Theories from Three-Manifolds},'' in {\em New
  Dualities of Supersymmetric Gauge Theories}, J.~Teschner, ed., pp.~339--373.
\newblock 2016.
\newblock
\href{http://www.arXiv.org/abs/1412.7129}{{\tt 1412.7129}}.
\newblock

\bibitem{Teschner:2014oja}
J.~Teschner, ``{Exact Results on ${\mathcal N}=2$ Supersymmetric Gauge
  Theories},'' in {\em New Dualities of Supersymmetric Gauge Theories},
  J.~Teschner, ed., pp.~1--30.
\newblock 2016.
\newblock
\href{http://www.arXiv.org/abs/1412.7145}{{\tt 1412.7145}}.
\newblock

\bibitem{Gang:2015wya}
D.~Gang, N.~Kim, M.~Romo, and M.~Yamazaki, ``{Aspects of Defects in 3d-3d
  Correspondence},''
\href{http://www.arXiv.org/abs/1510.05011}{{\tt 1510.05011}}.

\bibitem{Witten:1988hf}
E.~Witten, ``{Quantum Field Theory and the Jones Polynomial},'' {\em
  Commun.Math.Phys.} {\bf 121} (1989)
351.

\bibitem{Witten:1989ip}
E.~Witten, ``{Quantization of {Chern-Simons} Gauge Theory With Complex Gauge
  Group},'' {\em Commun. Math. Phys.} {\bf 137} (1991)
29--66.

\bibitem{Gukov:2003na}
S.~Gukov, ``{Three-dimensional quantum gravity, Chern-Simons theory, and the A
  polynomial},'' {\em Commun. Math. Phys.} {\bf 255} (2005) 577--627,
  \href{http://www.arXiv.org/abs/hep-th/0306165}{{\tt hep-th/0306165}}.

\bibitem{Witten:2010cx}
E.~Witten, ``{Analytic Continuation Of Chern-Simons Theory},''
\href{http://www.arXiv.org/abs/1001.2933}{{\tt 1001.2933}}.

\bibitem{Axelrod:1989xt}
S.~Axelrod, S.~Della~Pietra, and E.~Witten, ``{Geometric quantization of
  Chern-Simons gauge theory},'' {\em J. Diff. Geom.} {\bf 33} (1991), no.~3,
787--902.

\bibitem{Elitzur:1989nr}
S.~Elitzur, G.~W. Moore, A.~Schwimmer, and N.~Seiberg, ``{Remarks on the
  Canonical Quantization of the Chern-Simons-Witten Theory},'' {\em Nucl.Phys.}
  {\bf B326} (1989) 108.

\bibitem{Terashima:2013fg}
Y.~Terashima and M.~Yamazaki, ``{3d N=2 Theories from Cluster Algebras},'' {\em
  PTEP} {\bf 023} (2014) B01,
\href{http://www.arXiv.org/abs/1301.5902}{{\tt 1301.5902}}.

\bibitem{2011SIGMA...7..102K}
R.~M. {Kashaev} and T.~{Nakanishi}, ``{Classical and Quantum Dilogarithm
  Identities},'' {\em SIGMA} {\bf 7} (Nov., 2011) 102,
  \href{http://www.arXiv.org/abs/1104.4630}{{\tt 1104.4630}}.

\bibitem{Beasley:2009mb}
C.~Beasley, ``{Localization for Wilson Loops in Chern-Simons Theory},'' {\em
  Adv.Theor.Math.Phys.} {\bf 17} (2013) 1--240,
\href{http://www.arXiv.org/abs/0911.2687}{{\tt 0911.2687}}.

\bibitem{BN95}
D.~Bar-Natan, ``{On the Vassiliev Knot Invariants},'' {\em Topology} {\bf 34}
  (1995) 423.

\bibitem{Woodhouse:1992de}
N.~M.~J. Woodhouse, {\em {Geometric quantization}}.
\newblock
1992.
\newblock

\bibitem{Dimofte:2013iv}
T.~Dimofte, M.~Gabella, and A.~B. Goncharov, ``{K-Decompositions and 3d Gauge
  Theories},''
\href{http://www.arXiv.org/abs/1301.0192}{{\tt 1301.0192}}.

\bibitem{Dimofte:2009yn}
T.~Dimofte, S.~Gukov, J.~Lenells, and D.~Zagier, ``{Exact Results for
  Perturbative Chern-Simons Theory with Complex Gauge Group},'' {\em Commun.
  Num. Theor. Phys.} {\bf 3} (2009) 363--443,
\href{http://www.arXiv.org/abs/0903.2472}{{\tt 0903.2472}}.

\bibitem{HitchinSelfDuality}
N.~J. Hitchin, ``The self-duality equations on a {R}iemann surface,'' {\em
  Proc. London Math. Soc. (3)} {\bf 55} (1987), no.~1, 59--126.

\bibitem{CooperApolynomial}
D.~Cooper, M.~Culler, H.~Gillet, D.~D. Long, and P.~B. Shalen, ``Plane curves
  associated to character varieties of {$3$}-manifolds,'' {\em Invent. Math.}
  {\bf 118} (1994), no.~1, 47--84.

\bibitem{Freed:2008jq}
D.~S. Freed, ``{Remarks on Chern-Simons Theory},''
\href{http://www.arXiv.org/abs/0808.2507}{{\tt 0808.2507}}.

\bibitem{Dimofte:2012qj}
T.~D. Dimofte and S.~Garoufalidis, ``{The Quantum content of the gluing
  equations},''
\href{http://www.arXiv.org/abs/1202.6268}{{\tt 1202.6268}}.

\bibitem{Dimofte:2015kkp}
T.~Dimofte and S.~Garoufalidis, ``{Quantum modularity and complex Chern-Simons
  theory},''
\href{http://www.arXiv.org/abs/1511.05628}{{\tt 1511.05628}}.

\bibitem{1996q.alg.....1025K}
R.~M. {Kashaev}, ``{The hyperbolic volume of knots from quantum dilogarithm},''
  in {\em eprint arXiv:q-alg/9601025}, p.~1025.
\newblock Jan., 1996.

\bibitem{MurakamiMurakami}
H.~Murakami and J.~Murakami, ``The colored {J}ones polynomials and the
  simplicial volume of a knot,'' {\em Acta Math.} {\bf 186} (2001), no.~1,
  85--104.

\bibitem{2002math......3119M}
H.~{Murakami}, J.~{Murakami}, M.~{Okamoto}, T.~{Takata}, and Y.~{Yokota},
  ``{Kashaev's conjecture and the Chern-Simons invariants of knots and
  links},'' {\em ArXiv Mathematics e-prints} (Mar., 2002)
  \href{http://www.arXiv.org/abs/math/0203119}{{\tt math/0203119}}.

\bibitem{2010arXiv1003.4808D}
T.~{Dimofte} and S.~{Gukov}, ``{Quantum Field Theory and the Volume
  Conjecture},'' {\em ArXiv e-prints} (Mar., 2010)
  \href{http://www.arXiv.org/abs/1003.4808}{{\tt 1003.4808}}.

\bibitem{Dijkgraaf:2010ur}
R.~Dijkgraaf, H.~Fuji, and M.~Manabe, ``{The Volume Conjecture, Perturbative
  Knot Invariants, and Recursion Relations for Topological Strings},'' {\em
  Nucl. Phys.} {\bf B849} (2011) 166--211,
  \href{http://www.arXiv.org/abs/1010.4542}{{\tt 1010.4542}}.

\bibitem{Borot:2012cw}
G.~Borot and B.~Eynard, ``{All-order asymptotics of hyperbolic knot invariants
  from non-perturbative topological recursion of A-polynomials},''
\href{http://www.arXiv.org/abs/1205.2261}{{\tt 1205.2261}}.

\bibitem{Dimofte:2011gm}
T.~Dimofte, ``{Quantum Riemann Surfaces in Chern-Simons Theory},'' {\em Adv.
  Theor. Math. Phys.} {\bf 17} (2013) 479--599,
\href{http://www.arXiv.org/abs/1102.4847}{{\tt 1102.4847}}.

\bibitem{HikamiHyperbolic1}
K.~Hikami, ``Hyperbolic structure arising from a knot invariant,'' {\em
  Internat. J. Modern Phys. A} {\bf 16} (2001), no.~19, 3309--3333.

\bibitem{HikamiHyperbolic2}
K.~Hikami, ``Hyperbolic structure arising from a knot invariant. {II}.
  {C}ompleteness,'' {\em Internat. J. Modern Phys. B} {\bf 16} (2002),
  no.~14-15, 1963--1970. Lattice statistics \& mathematical physics, 2001
  (Tianjin).

\bibitem{2007JGP}
K.~{Hikami}, ``{Generalized volume conjecture and the A-polynomials: The
  Neumann Zagier potential function as a classical limit of the partition
  function},'' {\em Journal of Geometry and Physics} {\bf 57} (Aug., 2007)
  1895--1940, \href{http://www.arXiv.org/abs/math/0604094}{{\tt math/0604094}}.

\bibitem{Andersen:2011bt}
J.~Ellegaard~Andersen and R.~Kashaev, ``{A TQFT from Quantum Teichm\"uller
  Theory},'' {\em Commun.Math.Phys.} {\bf 330} (2014) 887--934,
\href{http://www.arXiv.org/abs/1109.6295}{{\tt 1109.6295}}.

\bibitem{Dimofte:2011py}
T.~Dimofte, D.~Gaiotto, and S.~Gukov, ``{3-Manifolds and 3d Indices},''
\href{http://www.arXiv.org/abs/1112.5179}{{\tt 1112.5179}}.

\bibitem{Dimofte:2014zga}
T.~Dimofte, ``{Complex Chern-Simons theory at level k via the 3d-3d
  correspondence},''
\href{http://www.arXiv.org/abs/1409.0857}{{\tt 1409.0857}}.

\bibitem{2014arXiv1409.1208E}
J.~{Ellegaard Andersen} and R.~{Kashaev}, ``{Complex Quantum Chern-Simons},''
  {\em ArXiv e-prints} (Sept., 2014)
  \href{http://www.arXiv.org/abs/1409.1208}{{\tt 1409.1208}}.

\bibitem{FockGoncharovHigher}
V.~Fock and A.~Goncharov, ``Moduli spaces of local systems and higher
  {T}eichm\"uller theory,'' {\em Publ. Math. Inst. Hautes \'Etudes Sci.}
  (2006), no.~103, 1--211.

\bibitem{Coman:2015lna}
I.~Coman, M.~Gabella, and J.~Teschner, ``{Line operators in theories of class
  $\mathcal{S}$, quantized moduli space of flat connections, and Toda field
  theory},''
\href{http://www.arXiv.org/abs/1505.05898}{{\tt 1505.05898}}.

\bibitem{FominZelevinsky}
S.~Fomin and A.~Zelevinsky, ``Cluster algebras. {I}. {F}oundations,'' {\em J.
  Amer. Math. Soc.} {\bf 15} (2002), no.~2, 497--529 (electronic).

\bibitem{Cordova:2013cea}
C.~Cordova and D.~L. Jafferis, ``{Complex Chern-Simons from M5-branes on the
  Squashed Three-Sphere},''
\href{http://www.arXiv.org/abs/1305.2891}{{\tt 1305.2891}}.

\bibitem{Hama:2011ea}
N.~Hama, K.~Hosomichi, and S.~Lee, ``{SUSY Gauge Theories on Squashed
  Three-Spheres},'' \href{http://www.arXiv.org/abs/1102.4716}{{\tt 1102.4716}}.

\bibitem{Imamura:2011wg}
Y.~Imamura and D.~Yokoyama, ``{N=2 supersymmetric theories on squashed
  three-sphere},'' {\em Phys.Rev.} {\bf D85} (2012) 025015,
\href{http://www.arXiv.org/abs/1109.4734}{{\tt 1109.4734}}.

\bibitem{Gueritaud}
F.~Gu{\'e}ritaud, ``On canonical triangulations of once-punctured torus bundles
  and two-bridge link complements,'' {\em Geom. Topol.} {\bf 10} (2006)
  1239--1284. With an appendix by David Futer.

\bibitem{Gang:2013sqa}
D.~Gang, E.~Koh, S.~Lee, and J.~Park, ``{Superconformal Index and 3d-3d
  Correspondence for Mapping Cylinder/Torus},''
\href{http://www.arXiv.org/abs/1305.0937}{{\tt 1305.0937}}.

\bibitem{2004math......6251P}
M.~{Polyak}, ``{Feynman diagrams for pedestrians and mathematicians},'' {\em
  ArXiv Mathematics e-prints} (June, 2004)
  \href{http://www.arXiv.org/abs/math/0406251}{{\tt math/0406251}}.

\bibitem{2013arXiv1310.3068K}
T.~{Kitayama} and Y.~{Terashima}, ``{Torsion functions on moduli spaces in view
  of the cluster algebra},'' {\em ArXiv e-prints} (Oct., 2013)
  \href{http://www.arXiv.org/abs/1310.3068}{{\tt 1310.3068}}.

\bibitem{Lee:2013ida}
S.~Lee and M.~Yamazaki, ``{3d Chern-Simons Theory from M5-branes},''
\href{http://www.arXiv.org/abs/1305.2429}{{\tt 1305.2429}}.

\bibitem{Yagi:2013fda}
J.~Yagi, ``{3d TQFT from 6d SCFT},'' {\em JHEP} {\bf 08} (2013) 017,
\href{http://www.arXiv.org/abs/1305.0291}{{\tt 1305.0291}}.

\bibitem{Chung:2014qpa}
H.-J. Chung, T.~Dimofte, S.~Gukov, and P.~Sulkowski, ``{3d-3d Correspondence
  Revisited},''
\href{http://www.arXiv.org/abs/1405.3663}{{\tt 1405.3663}}.

\bibitem{2014arXiv1412.4711F}
E.~{Falbel}, A.~{Guilloux}, P.-V. {Koseleff}, F.~{Rouillier}, and
  M.~{Thistlethwaite}, ``{Character varieties for SL(3,C): the figure eight
  knot},'' {\em ArXiv e-prints} (Dec., 2014)
  \href{http://www.arXiv.org/abs/1412.4711}{{\tt 1412.4711}}.

\bibitem{2015arXiv150504451H}
M.~{Heusener}, V.~{Munoz}, and J.~{Porti}, ``{The SL(3,C)-character variety of
  the figure eight knot},'' {\em ArXiv e-prints} (May, 2015)
  \href{http://www.arXiv.org/abs/1505.04451}{{\tt 1505.04451}}.

\bibitem{Gaiotto:2008ak}
D.~Gaiotto and E.~Witten, ``{S-Duality of Boundary Conditions In N=4 Super
  Yang-Mills Theory},'' {\em Adv.Theor.Math.Phys.} {\bf 13} (2009) 721,
\href{http://www.arXiv.org/abs/0807.3720}{{\tt 0807.3720}}.

\bibitem{Xie:2012dw}
D.~Xie, ``{Network, Cluster coordinates and N=2 theory I},''
\href{http://www.arXiv.org/abs/1203.4573}{{\tt 1203.4573}}.

\bibitem{Gang:2015bwa}
D.~Gang, N.~Kim, M.~Romo, and M.~Yamazaki, ``{Taming Supersymmetric Defects in
  3d-3d Correspondence},''
\href{http://www.arXiv.org/abs/1510.03884}{{\tt 1510.03884}}.

\bibitem{FaddeevKashaevQuantum}
L.~D. Faddeev and R.~M. Kashaev, ``Quantum dilogarithm,'' {\em Modern Phys.
  Lett. A} {\bf 9} (1994), no.~5, 427--434.

\bibitem{Faddeev95}
L.~D. Faddeev, ``Discrete {H}eisenberg-{W}eyl group and modular group,'' {\em
  Lett. Math. Phys.} {\bf 34} (1995), no.~3, 249--254,
  \href{http://www.arXiv.org/abs/hep-th/9504111}{{\tt hep-th/9504111}}.

\bibitem{FaddeevModular}
L.~Faddeev, ``Modular double of a quantum group,'' in {\em Conf\'erence
  {M}osh\'e {F}lato 1999, {V}ol. {I} ({D}ijon)}, vol.~21 of {\em Math. Phys.
  Stud.}, pp.~149--156.
\newblock Kluwer Acad. Publ., Dordrecht, 2000.

\bibitem{Faddeev:2000if}
L.~D. Faddeev, R.~M. Kashaev, and A.~Y. Volkov, ``{Strongly coupled quantum
  discrete Liouville theory. I: Algebraic approach and duality},'' {\em Commun.
  Math. Phys.} {\bf 219} (2001) 199--219,
\href{http://www.arXiv.org/abs/hep-th/0006156}{{\tt hep-th/0006156}}.

\end{thebibliography}\endgroup





\end{document}